\documentclass[fleqn,usenatbib]{mnras/mnras}

\usepackage{newtxtext,newtxmath}

\usepackage[T1]{fontenc}

\DeclareRobustCommand{\VAN}[3]{#2}
\let\VANthebibliography\thebibliography
\def\thebibliography{\DeclareRobustCommand{\VAN}[3]{##3}\VANthebibliography}

\usepackage{graphicx}	
\usepackage{amsmath}	
\usepackage[utf8]{inputenc}
\usepackage{siunitx}
\usepackage{multirow}

\newif\ifptitle 
\newif\ifpnumber

\usepackage[dvipsnames,svgnames,x11names]{xcolor}

\usepackage{soul}

\title[Identifying Galaxy Cluster Mergers with Deep Neural Networks using Idealized Compton-y and X-ray maps]{Identifying Galaxy Cluster Mergers with Deep Neural Networks using Idealized Compton-y and X-ray maps}

\author[Arendt, A. et al.]{Ashleigh R. Arendt,$^{1}$\thanks{E-mail: arendtash@hotmail.com}
Yvette C. Perrott,$^{1}$
Ana Contreras-Santos,$^{2}$
Daniel de Andres,$^{3}$
Weiguang Cui,$^{4}$
\newauthor
Douglas Rennehan$^{5}$ 
\\
$^{1}$ School of Chemical and Physical Sciences, Victoria University of Wellington, PO Box 600, Wellington 6140, New Zealand \\
$^{2}$ Departamento de Física Teórica, Módulo 15, Facultad de Ciencias, Universidad Autónoma de Madrid, 28049 Madrid, Spain \\
$^{3}$ Departamento de Física Teórica and CIAFF, Módulo 8, Universidad Autónoma de Madrid, 28049 Madrid, Spain \\
$^{4}$ Institute for Astronomy, University of Edinburgh, Royal Observatory, Edinburgh EH9 3HJ, UK \\
$^{5}$ Center for Computational Astrophysics, Flatiron Institute, 162 Fifth Avenue, New York, NY, 10010, USA \\
}

\date{Accepted XXX. Received YYY; in original form ZZZ}

\pubyear{2023}

\begin{document}
\label{firstpage}
\pagerange{\pageref{firstpage}--\pageref{lastpage}}
\maketitle

\begin{abstract}
We present a novel approach to identify galaxy clusters that are undergoing a merger using a deep learning approach. This paper uses massive galaxy clusters spanning $0 \leq z \leq 2$ from \textsc{The Three Hundred} project, a suite of hydrodynamic re-simulations of 324 large galaxy clusters. Mock, idealised Compton-{\it y} and X-ray maps were constructed for the sample, capturing them out to a radius of $2R_{200}$. The idealised nature of these maps mean they do not consider observational effects such as foreground or background astrophysical objects, any spatial resolution limits or restriction on X-ray energy bands. Half of the maps belong to a merging population as defined by a mass increase $\Delta${\it M/M} $\geq$ 0.75, and the other half serve as a control, relaxed population. We employ a convolutional neural network architecture and train the model to classify clusters into one of the groups. A best-performing model was able to correctly distinguish between the two populations with a balanced accuracy (BA) and recall of 0.77, ROC-AUC of 0.85, PR-AUC of 0.55 and $F_{1}$ score of 0.53. Using a multichannel model relative to a single channel model, we obtain a 3\% improvement in BA score, and a 6\% improvement in $F_{1}$ score. We use a saliency interpretation approach to discern the regions most important to each classification decision. By analysing radially binned saliency values we find a preference to utilise regions out to larger distances for mergers with respect to non-mergers, greater than $\sim1.2 R_{200}$ and $\sim0.7 R_{200}$ for SZ and X-ray respectively.
\end{abstract}




\section{Introduction}
\label{sec:Intro}

Galaxy clusters are the largest-scale virialised bodies in our Universe and can be used to probe the formation and evolution of structure along with constraining cosmological parameters. Specifically, the abundance of clusters as a function of their halo mass can constrain the late-Universe matter density parameter $\Omega_{m}$ and matter fluctuation amplitude $\sigma_{8}$ \citep{Aghanim2020}. Cluster-cluster mergers are one of the main drivers of cluster instability and can lead to large uncertainties in standard methods of estimating cluster mass. One example of this is when obtaining mass using the hydrostatic assumption \citep[e.g.][]{Gianfagna2021,Gianfagna2023}. Therefore, it is imperative to determine whether a cluster is undergoing a merger when using cluster observations to constrain cosmological parameters. 

Clusters accumulate mass by growing hierarchically through either gradual accretion of individual and smaller groups of galaxies, or merger events with larger groups and galaxy clusters \citep{Kravtsov2012}. With such large potential wells, cluster merger events release immense amounts of energy through mechanisms that are not well understood. \citet{Mann_2012} and \citet{Lokas2023} are among the studies that have previously aimed to identify cluster mergers, both in observation and simulation respectively. This identification process facilitates the study of the unique physics at play in these environments. As larger statistical samples of merger events become available, works that analyse the nature and evolution of a small number of mergers (e.g. \citealt{Wilber2019}) can be extended. Quantifying the disturbance of a cluster is non-trivial and has resulted in varying definitions across the literature for relaxed and disturbed states \citep{De_Luca_2021}. This is true with both theoretical probes \citep[e.g.][]{Cui2017,Haggar_2020,Li_2021, Zhang_2022} and morphological indicators, which have been proposed and evaluated across multi-wavelength observations \citep[e.g.][]{Capalbo_2020, Yuan_2020, Zenteno_2020}. Merging processes contribute greatly to the dynamical state of a cluster; better understanding and identification of mergers can therefore help to better describe and constrain their dynamical state.   

An intracluster medium of hot gas pervades each cluster and provides an observational probe to study its properties. Different wavelengths trace different processes within the gas; X-ray observations primarily track thermal Bremsstrahlung emission from the hot gas, whereas millimetre observations show the scattering of cosmic microwave background (CMB) photons by high energy electrons through the Sunyaev-Zel’dovich (SZ) effect.

In this paper, we provide a new method for classifying a cluster as a merging candidate, with the use of machine learning. Machine learning enables the discovery of complex, non-linear relationships between observable and physical properties in a data-driven approach. Previously, machine learning has been applied to astrophysical contexts from solar wind \citep{Wrench_2022} to exoplanet detection \citep{Malik_2021}, supernova classification \citep{Lochner_2016} and painting N-body simulations with baryonic properties \citep{de_andres_2022_nbody}. It has become an increasingly used tool in the field of galaxy clusters, mainly to reduce the scatter in mass estimates compared with standard scaling relations. This has been achieved using X-ray \citep{Green2019, Ntampaka2019, Ho2023, Krippendorf2023}, SZ \citep{Gupta2020, Gupta2021, deandres22_cnn, Wadekar2022, Ferragamo2023}, optical \citep{Ntampaka2015, Ntampaka2016, Ho2019,  Ho2021, Ho2022, Ramanah2020a, Ramanah2020b, Lin_2022} and multi-wavelength studies \citep{Armitage2018, Cohn2020, Yan2020}. Optical images have also been combined with machine learning techniques to identify merging galaxy systems in galaxy redshift catalogues \citep{delosrios16}. \citet{Su2020} implement a classification algorithm using mock {\it Chandra} X-ray images to categorise clusters based on the properties of the cluster core. Whereas, \citet{Li_2022} extract dynamical features from optical, SZ and X-ray information to determine which morphological and dynamical indicators best describe the dynamical state of a cluster. 

In this paper, we extend the range of machine learning applications by implementing statistical techniques to identify galaxy cluster mergers in simulation data. We make use of data from \textsc{The Three Hundred} catalogue \citep{Cui_2018}, a suite of 324 clusters re-simulated to include full particle hydrodynamics. Images are generated simulating both X-ray and SZ observations, and these idealised images are fed into a network to output a merger probability. Throughout this paper, as in \textsc{The Three Hundred} simulations, we assume Planck 2015 cosmology \citep{Ade2016} with $h = H_{0} / 100 $ km s$^{-1}$ Mpc$^{-1}$ = 0.678, $\Omega_\mathrm{M} = 0.307$, $\Omega_\mathrm{B} = 0.048$, $\Omega_\mathrm{\Lambda} = 0.693$, $\sigma_{8} = 0.823$, $n_\mathrm{s} = 0.96$. All $M_{200}$ values refer to the mass included within $R_{200}$, the radius at which the density of the material enclosed $M({<}r) / (4\pi r^{3}/3)$ is equal to $200\rho_\mathrm{crit}$, or 200 times the critical density of the Universe at that redshift.

The paper is organised as follows.  In Section \ref{sec:Data} the data sources are explained and in Section \ref{sec:Identifying Mergers} descriptions are given for how mergers are defined and identified.  In Section \ref{sec:Methods} machine learning methods are described, followed by an analysis of the model performance in Section \ref{sec:Results}, interpretation and discussion in Section \ref{sec:Discussion}, and conclusion in Section \ref{sec:Conclusion}.

\section{Data Sources}
\label{sec:Data}
\subsection{The Three Hundred}
\label{subsec:TheThreeHundred}
The 324 clusters in \textsc{The Three Hundred} project are a selection of the most massive clusters taken from the MultiDark Planck 2 dark matter (DM) only simulation (MDPL2, \citealt{Klypin2016}), identified at $z = 0$. MDPL2 uses {\it Planck} 2015 cosmology \citep{Ade2016} to simulate a periodic cube of comoving length 1 $h^\mathrm{-1}$ Gpc containing $3840^\mathrm{3}$ DM particles of individual mass $1.5 \cdot 10^\mathrm{9} h^\mathrm{-1}$ \(\textup{M}_\odot\). These clusters are identified using the \textsc{rockstar} halo finder \citep{Behroozi_2012}. A high-resolution spherical region of 15 $h^\mathrm{-1}$ Mpc surrounding each cluster is resimulated with different baryonic physics models: \textsc{Gadget-MUSIC} \citep{Sembolini2013}, \textsc{Gadget-X}\citep{Rasia2015}, and \textsc{GIZMO-SIMBA}\citep{Dave2019,Cui2022}. In this paper, we use the data generated by the \textsc{Gadget-X} code, which uses a modern Smooth-Particle-Hydrodynamics (SPH) solver to simulate the clusters from initial conditions. The output of the re-simulations includes 129 snapshots for each of the respective 324 clusters, tracking the clusters through time as they evolved from $z = 16.98$ to $z = 0$. At $z = 0$ the clusters range from masses $M_{200} = 6.4 \cdot 10^\mathrm{14} h^\mathrm{-1}$ \(\textup{M}_\odot\) to $M_{200} = 2.65 \cdot 10^\mathrm{15} h^\mathrm{-1}$ \(\textup{M}_\odot\). Refer to \citet{Cui_2018} for more details about the code and data sets behind the cluster simulations. 


Halo properties for each region were extracted using the \textsc{ahf}\footnote{http://popia.ft.uam.es/AHF/} halo finder \citep{Knollmann_2009}, which includes both gas and stars in the halo finding process and identifies structure through overdensities in an adaptively smoothed density field \citep[see e.g.][for more details on halo finders]{Knebe_2011}. $R_\mathrm{200}$ is calculated for each halo; subhaloes are identified as those that reside within the central $R_\mathrm{200}$ of another halo, termed the `host halo'. 
To extract merger histories for each halo, tracing objects across redshift, the \textsc{ahf} package contains \textsc{mergertree}. This tool takes each halo identified at $z = 0$ and follows it backwards in time, employing a merit function to identify all progenitors in the previous snapshot in order of size. The merit function is given as: $\mathcal{M} = N^\mathrm{2}_\mathrm{AB} / (N_\mathrm{A}N_\mathrm{B})$ for a halo A with a main progenitor halo B, where B maximises this function. $N_\mathrm{A}$ and $N_\mathrm{B}$ are the number of particles within halo A and B respectively, and $N_\mathrm{AB}$ is the number within both. The tool can also skip snapshots, allowing for a continuation of the tree in the case where no suitable progenitor is found in the preceding snapshot. 
See \citet{Srisawat_2013} for more information on \textsc{mergertree} and other tree-finding algorithms.

\subsection{SZ image generation}
The SZ effect can be split into two main effects: thermal (tSZ) and kinetic (kSZ); see \citet{Mroczkowski2018} for a review. Here, we focus solely on tSZ as the effect that causes the greatest distortion in the CMB signal. The tSZ effect is driven by the motions of thermal electrons which can be quantified as the Compton-{\it y} parameter according to \noindent Equation \ref{eqn:comptony},

\begin{equation}
\label{eqn:comptony}
y =  \frac {\sigma_\mathrm{T}k_\mathrm{B}}{m_\mathrm{e}c^\mathrm{2}} \int n_\mathrm{e}T_\mathrm{e}\,dl \,,
\end{equation}

where $k_\mathrm{B}$ is the Boltzmann constant, $\sigma_\mathrm{T}$ is the Thomson cross-section, $m_\mathrm{e}$ is the electron rest mass, $c$ is the speed of light, $n_\mathrm{e}$ and $T_\mathrm{e}$ are the electron number density and electron temperature respectively, integrated along the line of sight $dl$.

To generate the tSZ mock maps we make use of the PyMSZ\footnote{https://github.com/weiguangcui/pymsz} package \citep{Cui_2018} 
which produces 2D projections of the Compton {\it y}-parameter for the regions centred at the galaxy clusters. This is achieved by substituting $dl$ by $dV/dA$, and  $n_\mathrm{e}$ as $N_\mathrm{e}/dV = N_\mathrm{e}/dA/dl $
resulting in Equation \ref{eqn:comptony_pymsz},

\begin{equation}
\label{eqn:comptony_pymsz}
y \simeq  \frac {\sigma_\mathrm{T}k_\mathrm{B}}{m_\mathrm{e}c^\mathrm{2}dA} \sum_{i} N_\mathrm{e,i}T_\mathrm{e,i}W(r, h_\mathrm{i}) \,,
\end{equation}
where $N_\mathrm{e}$ is the number of electrons in a gas particle, $V$ is the gas volume, $dA$ is the projected area perpendicular to $dl$ and $W(r, h_\mathrm{i})$ is the projected SPH kernel with smoothing length $h_\mathrm{i}$.

We need a consistent image size to input into our model, so we set the number of pixels per side, $N_\mathrm{pix}$, to 640. We also want the same field of view for each cluster, to understand which regions are important in determining a merger. To do this, we adjust the angular resolution of the mock images (or the distances to the clusters) to make sure that the image always covers a radius of $2R_\mathrm{200}$ of the `observed' cluster, i.e. $R_\mathrm{200} = 160$px. The physical properties of the cluster are kept to its simulated redshift. Each cluster is rotated along 29 different lines of sight before a 2D projected {\it y}-map is generated. The 29 line of sight projections are determined by equally dividing the surface area of a sphere without including reflections / mirrored projections. The number 29 was chosen to balance between the number of maps in the training sample and to avoid identical maps. Note that for disturbed clusters, all these 29 projections are distinct.

\subsection{X-ray image generation} X-ray emission can be attributed to different emission processes within the collisional plasma that makes up the intracluster medium. The predominant mechanism is thermal Bremsstrahlung, whereby free-moving electrons are deflected by nuclei releasing photons as a continuum process. Heavier elements also contribute through line emission processes as electrons move between quantum states. In this study, we generate X-ray images in the form of bolometric luminosity maps constructed in the energy range 0.1 -- 15.0 keV. This serves as an idealistic wavelength range -- the simplest case -- to test its performance. This is wider than we see in typical X-ray surveys, e.g. 0.5 -- 10.0 keV for the eROSITA full energy range, however, this can be easily recalibrated when using the analysis method for real cluster images. To create these maps, the simulated particles within each cluster halo are projected onto a 2D grid using a nearest grid point algorithm, creating a grid of side length 4$R_\mathrm{200}$, the same as the case for SZ mock images.

Gas temperatures are provided by the \textsc{Gadget-X} simulation from the internal energies and gas masses, assuming an ideal monotomic gas with isentropic expansion factor $\gamma  = 5/3$. Modelled data is restricted to the hot, low-density gas and power curves are constructed using the \textsc{xraylum}\footnote{https://github.com/rennehan/xraylum} python package. This code builds on top of \textsc{pyatomdb}\footnote{https://github.com/AtomDB/pyatomdb} and runs \textsc{apec} \citep{Foster_2012} for our chosen energy range to calculate the cooling curves for Bremsstrahlung, H, He, C, N, Si, and Fe emissions, assuming a hydrogen fraction of 0.76. The individual power contributions are then extracted for each gas location using the temperature, linearly combined for each element and multiplied by a gas factor representing the number of particles at that location. For each of the 29 projection axes, the gas coordinates are rotated and the total power is pooled along the line of sight and binned into an image with $N_\mathrm{pix} = 640$, where the pixel count represents the X-ray luminosity. To speed up image generation time, SPH particle smoothing is not employed as it is for the SZ map creation, however, at the resolution of the images input to the model this has minimal impact. 

\subsection{Idealised image generation} 
\begin{figure}
	\includegraphics[width=\columnwidth]{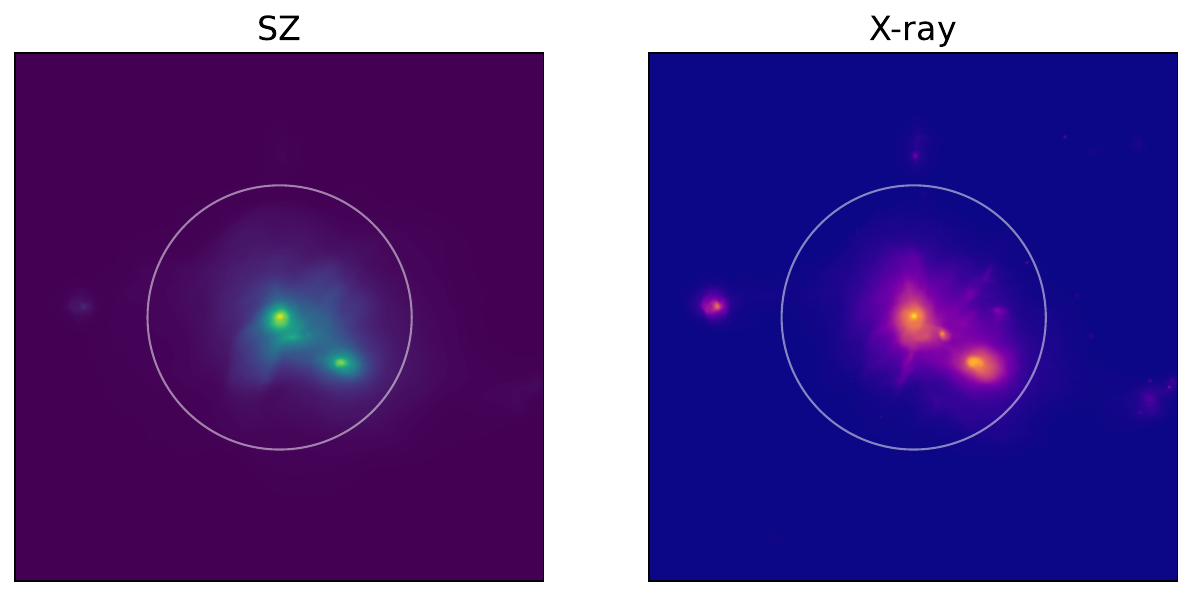}
    \caption{Example input image of merging cluster in both SZ and X-ray filters. White circle represents the $R_{200}$ radius.}
    \label{fig:example_image}
\end{figure}

The images generated in this study are idealised and serve as a proof of concept for identifying merging clusters using deep learning, without any observational effects such as resolution, instrumental noise or contaminant sources. An example projection image of a merging cluster is shown in Fig.\ref{fig:example_image} generated for both SZ and X-ray. We adjust the angular resolution scale of simulated clusters such that we would be able to observe them similarly regardless of their redshift. In reality, many of the higher redshift clusters included in this sample would produce very faint structures at resolutions and sensitivities available to modern telescopes. The inclusion of a range of redshifts also means there are some physical differences in the cluster properties due to their evolution history, such as mass and star formation rate, which, however, precisely follow the simulation results. The aim is to stage this approach, firstly assessing the performance of the idealised images, then in later work producing images at a realistic resolution and sensitivity to mimick current and upcoming imaging surveys.

By including rotations along 29 different lines of sight we create a more generalisable model, increasing the statistical sample of the images the model is `seeing'. However, it is worth noting that the underlying mass and redshift distribution of the clusters is \textit{not} being augmented, so the diversity of these properties that the model sees is inherently limited by the number of merging samples that exist in \textsc{The Three Hundred} data sample. Nevertheless, using different rotations takes the projection effect into account, i.e. a true merger can not be identified if the two main subclusters are projected onto the same line of sight. 

\section{Identifying mergers}
\label{sec:Identifying Mergers}

Merging galaxy clusters are complicated systems that can broadly be defined by three evolutionary phases: a pre-merger phase, where the intracluster medium gas from the two progenitors begins to interact and become dynamically disturbed; a main merging phase where the majority of mass accretion occurs and the cluster cores are interacting; and, a post-merger phase where the intracluster medium relaxes and energy is dissipated through turbulence and shocks (see \citet{Wilber2019} for real examples and descriptions of these phases). The duration and intensity of each phase depends on the unique physics of the merging systems: specifically, the mass ratios of the progenitors and the speed at which they collide. To identify merging halos and their phases at a given snapshot we follow the same logic as outlined in \citet{Contreras2022} (hereafter\defcitealias{Contreras2022}{CS22}\citetalias{Contreras2022}). The process is summarised here, but for more details please refer to CS22.

\subsection{Mass accretion history}
\label{subsec:MAH}
To build a set of merging clusters, we focus on the halos located in the central regions of the simulation at $z = 0$, solely tracing them back along their main branch using merger trees to build out their mass accretion history (MAH). Therefore, we limit the merger sample set to include the most massive progenitors of the central halo at each redshift. We use the MAH of each cluster to identify phases in which the cluster gained significant mass over short periods of time. Rapid accretion can be attributed to a merger event, and previous studies have aimed to quantify the time over which this mass must be accumulated to distinguish it from general accretion (e.g. \citealt{Cohn_2005, Wetzel_2007}). \citetalias{Contreras2022} combines previous work to create a logical definition of a merger for the data collected in \textsc{The Three Hundred}. Here, they look for a significant increase in mass over half the dynamical time of the cluster ($t_\mathrm{d}$) where $t_\mathrm{d}$ is defined as the crossing time, or the time it takes for a particle to complete a significant fraction of its orbit. We can write the dynamical time as
\begin{equation}
\label{eqn:dynamical_time}
t_\mathrm{d} = \sqrt{\frac{3}{4\pi}\frac{1}{200G\rho_\mathrm{crit}}}
\end{equation}
where $\rho_\mathrm{crit}$  is independent of the cluster itself, and depends only on the redshift at which the cluster is observed for a given cosmology. $t_\mathrm{d}$ therefore serves as a mass- and size-independent timescale.

We adopt the same timescale as in \citetalias{Contreras2022} but a different characteristic mass increase over that period. In our case, we define the mass increase required for a merger as at least 75\% of the initial mass, i.e.
\begin{equation}
\label{eqn:mass_increase}
\frac{\Delta M}{M} = \frac{M_\mathrm{f} - M_\mathrm{i}}{M_\mathrm{i}} \ge 0.75
\end{equation}
where $M_\mathrm{i}$ and $M_\mathrm{f}$ are the initial and final cluster mass values for the time interval $t_\mathrm{f} - t_\mathrm{i} \leq t_\mathrm{d} / 2$. In contrast, \citetalias{Contreras2022} uses a mass increase of at least 100\%. 
An interval of two snapshots always creates a time delay of $\lessapprox t_\mathrm{d} / 2$; we use this interval when measuring the fractional change in mass. We note that using two snapshots is an upper limit and if the mass increase is instead obtained over two successive snapshots, with an interval of one snapshot, then it is also identified as a merger. 

The lower mass increase threshold compared to \citetalias{Contreras2022} is reasoned due to the differing goals of each study. This study aims to create a model that learns the characteristics of a merger event, including both major and minor merger types. Decreasing the threshold from 100\% to 75\% allows for more of the minor merger population to be included in the sample, both increasing the training sample size and representing a wider range of merger scenarios (see Section \ref{subsec:MergerSampleCreation} for more details). 

Due to the definition of $M_\mathrm{200}$ being intrinsically linked to $\rho_\mathrm{crit}$ which varies across time, the halo mass undergoes pseudo-evolution. However, considering that mass accumulation associated with pseudo-evolution is more gradual than for merger events, and that this effect is more significant for galaxy-sized halos than cluster scales, \citetalias{Contreras2022} deem this effect to be negligible.

\subsection{Dynamical state evolution}
We combine the mass-based indicator with dynamical state indicators to fully describe a merger event. Merger events significantly disturb clusters; we expect that as a cluster undergoes a merger it will increasingly deviate from spherical symmetry until reaching a point of maximum disturbance, before then relaxing back to a stable state. There are a plethora of morphological indicators that can be combined to discern the dynamical state of a cluster. Here we follow the criteria for relaxation using a set of parameters developed in \citet{Cui2017} and later adopted in \citet{Haggar_2020,  Contreras2022, Zhang_2022}. The parameters considered consist of:
\begin{itemize}
    \item \textbf{centre of mass offset,} $\boldsymbol{\Delta_{r}}$: the offset of the cluster centre of mass from the density peak of the cluster halo obtained using the \textsc{ahf}, as a fraction of the cluster radius, $R_\mathrm{200}$;
    \item \textbf{subhalo mass fraction,} $\boldsymbol{f_{s}}$: the fraction of the mass of the cluster contained in subhalos; and 
    \item \textbf{the virial ratio,} $\boldsymbol{\eta}$: how closely the cluster obeys the virial theorem, defined as $ \eta = (2T - E_{s}) / |W|$, where T is the total kinetic energy, $E_{s}$ is the energy from surface pressure, and W is the total potential energy. 
    
\end{itemize}
These three indicators are taken together in \citet{Haggar_2020} to form a `relaxation' parameter, defined as in Equation \ref{eqn:relaxation_param},
\begin{equation}
\label{eqn:relaxation_param}
\chi_{DS} = \sqrt{\frac{3}{
\left( \frac{\Delta_{r}}{0.04}\right)^2 +
\left(\frac{f_{s}}{0.1}\right)^2 +
\left(\frac{|1-\eta|}{0.15}\right)^2 
}}
\end{equation}
The relaxation parameter is used to obtain a continuous measure of dynamical state, whereby for a cluster to be described as relaxed, we expect $\Delta_{r}$ and $f_{s}$ to be minimised and $\eta \to 1$ \citep{Haggar_2020}. By this logic, any clusters with $\chi_{DS} \geq 1$ are thought to be dynamically relaxed, and a drop in the $\chi_{DS}$ parameter is associated with a disturbance of the cluster.
Each parameter captures different attributes of cluster relaxation, as described in \citetalias{Contreras2022}.  $\Delta_{r}$ appears to contribute most to the general evolution of $\chi_{DS}$ across redshift, with $f_{s}$ following a similar trend. $\eta$ on the other hand, is responsible for dramatic changes in $\chi_{DS}$, rather than the general shape of the $\chi_{DS} (z)$ curve. By incorporating the differing information captured by each metric we create a more robust measure of the dynamical state.

\subsection{Characteristic merger times}
\label{subsec:MergerTimescales}
\begin{figure*}
	\includegraphics[width=\textwidth]{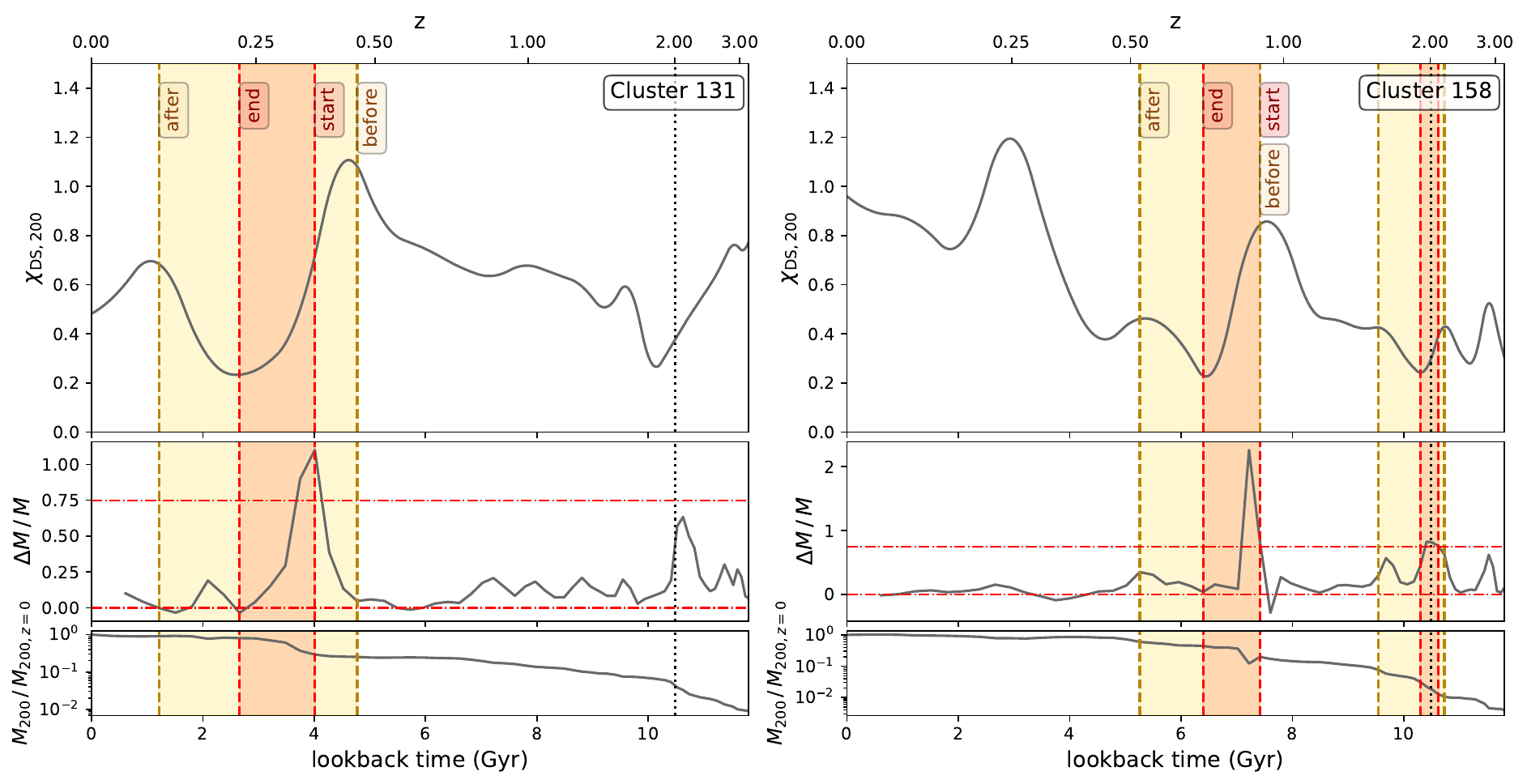}
    \caption{Mass accretion and relaxation history of two example clusters from our merging clusters data set, located at regions 131 and 158. The top panel describes $\chi_{DS}(t)$, the middle panel tracks the mass increase as described by \ref{eqn:mass_increase}, with a horizontal line marking the mass increase threshold for our merger definition, and the bottom panel is the ratio of the mass of the cluster ($M_{200}$) throughout time with respect to its $M_{200}$ value at $z = 0$. 
    Red vertical dashed lines indicate the start ($z_\mathrm{start}$) and end ($z_\mathrm{end}$) of the merger event, from right to left. Yellow dashed lines mark the corresponding dynamically disturbed periods pre- (from $z_\mathrm{before}$) and post-merger (until $z_\mathrm{after}$). The black vertical dashed line indicates our $z = 2$ threshold for the merging cluster sample.}
    \label{fig:MAH_XDS_evolution}
\end{figure*}

In order to categorise the clusters in our catalogue into merging phases we combine information on the cluster MAH and relaxation parameter across time. This is illustrated in Fig.~\ref{fig:MAH_XDS_evolution}, which follows the evolution of two representative clusters, located at simulated regions 131 and 158. The figure has three panels describing the evolution of relaxation parameter, relative mass change and MAH across time, respectively. The top panel shows the evolution of $\chi_\mathrm{DS,200}(t)$, the relaxation parameter calculated within $R_{200}$ (hereby $\chi_\mathrm{DS}$). This is calculated by taking a moving average of three successive snapshots and interpolating to filter the scatter in individual measurements. The middle panel shows the evolution of  $(\Delta M / M)(t)$, with a red horizontal line marking the condition for merging at $\Delta M / M = 0.75$. The bottom panel also represents the MAH, but traces $M_\mathrm{200}(t)/M_\mathrm{200,z=0}$, showing the cumulative gain in mass across the history of the cluster. We clearly see a relationship between sharp increases in cluster mass and declining values of the relaxation parameter. 

With reference to Fig.~\ref{fig:MAH_XDS_evolution}, \citetalias{Contreras2022} define four characteristic times for a merger event which we adopt here:
\begin{enumerate}
    \item $\boldsymbol{z_\mathrm{before}}$: the time just before a merger event takes place, when the cluster is dynamically relaxed, reflecting the point at which the clusters begin to interact. This is located by finding the snapshot corresponding to a maximum in the relaxation parameter before the merger begins;
    \item $\boldsymbol{z_\mathrm{start}}$: the onset of the merger, or point at which our merging criteria are met and $\Delta M / M \geq 0.75$. This is identified solely using the $(\Delta M / M)(t)$ curve at the redshift that marks the point in which our merger condition is met as outlined in Section \ref{subsec:MAH};
    \item $\boldsymbol{z_\mathrm{end}}$: the end of the merger itself, where disturbance from the merger event ends and the cluster begins to relax back to an equilibrium state. We combine both MAH and $\chi_\mathrm{DS}$ to define this point as the next minimum in $\chi_\mathrm{DS}$ after $z_\mathrm{start}$; and
    \item $\boldsymbol{z_\mathrm{after}}$: the time at which the cluster reaches a new relaxed phase, where $\chi_\mathrm{DS}$ reaches a new maximum after the merger.
\end{enumerate}

Fig.~\ref{fig:MAH_XDS_evolution} clearly highlights the three phases of merger evolution as follows:
\begin{enumerate}
    \item \textbf{Pre-merger phase}: Yellow shaded region from $z_\mathrm{before}$ to $z_\mathrm{start}$;
    \item \textbf{Merging phase}: Red shaded region from $z_\mathrm{start}$ to $z_\mathrm{end}$;
    \item \textbf{Post-merger phase}: Yellow shaded region from $z_\mathrm{end}$ to $z_\mathrm{after}$.
\end{enumerate}
In this study, we form a binary classification problem, and as such we group the clusters from each phase into one `merging' group. In terms of Fig.~\ref{fig:MAH_XDS_evolution} any snapshot taken at $z \leq 2$ that occurs within the shaded region will be included in the merging class, and the remaining snapshots can be sampled to create the non-merging class.

Since we are measuring these timescales in a discrete fashion through snapshots, there are cases where two of the characteristic times coincide; an example of this is shown in cluster 158 in Fig.~\ref{fig:MAH_XDS_evolution}, whereby $z_\mathrm{start}$ and $z_\mathrm{before}$ are located at the same redshift. This can be observed when two clusters merge rapidly and therefore their intracluster media interact and are disturbed on a similar timescale to the actual mass accretion. This time resolution is not an issue in this study as we are not distinguishing between the different merger phases when making classifications, but should be considered for future potential multi-class models.

\subsection{Merger sample creation}
\label{subsec:MergerSampleCreation}
When creating our merger sample we have to consider a balance of data quality, maximising the likelihood of an event being a merger, and data representation. We previously defined a merger event as one in which the mass of an individual halo increases by 75\% over half the dynamical time of the cluster, $\Delta${\it M/M} $\geq$ 0.75, in contrast with \citetalias{Contreras2022} (which uses a higher threshold of $\Delta${\it M/M} $\geq$ 1). We could have set this threshold at an even lower value. These values are somewhat arbitrary as merging systems take a range of forms. This section will explain the reasoning behind the thresholds chosen to select our merging sample. 

One factor in defining our sample is the number of particles that reside in the halo at $z_\mathrm{start}$, $N_\mathrm{part}$. We adopt $N_\mathrm{part} = 10,000$ in this study, following the analysis by \citetalias{Contreras2022}. This relatively high $N_\mathrm{part}$ value serves to reduce any possible uncertainties in the simulated halo morphology. It also shifts the distribution of clusters that fit the criteria to objects that we may expect to see at lower redshift. This better represents our target input prediction group for observation, since with current telescope resolution limits we are most likely to be able to apply the model to lower redshift populations.

To understand our thresholds better we explain some common conventions for categorising mergers. Cluster merger events are classified in literature based on the mass ratio of the two progenitors (${M_\mathrm{2}/M_\mathrm{1}}$), where $M_\mathrm{1}$ is the mass of the main branch merging halo, and $M_\mathrm{2}$ is the merging halo, with both masses measured at time $z_\mathrm{start}$ \citep{Planelles_2009, Chen_2019}. Minor mergers are those where 0.1 $\leq$ $|M_\mathrm{2}/M_\mathrm{1}| \leq 0.3 $, and major events are those with $|M_\mathrm{2}/M_\mathrm{1}|$ $\geq$ 0.3. We calculate this ratio for each merger event at $z_\mathrm{start}$, tracing the main progenitors using \textsc{mergertree} and merit functions described in Section \ref{subsec:TheThreeHundred}. The first cut we make is to exclude events that do not fit into either major or minor merger population, i.e. those where $|M_\mathrm{2}/M_\mathrm{1}|\leq 0.1$. The small mass ratio of the main branch halo and main progenitor implies that for the required increase in mass, the cluster likely merged with multiple substructures. In the case of multiple merging systems, the dynamics can be quite different to a classical binary merger; we exclude such cases from this study leaving for future work.

\begin{figure}
	\includegraphics[width=\columnwidth]{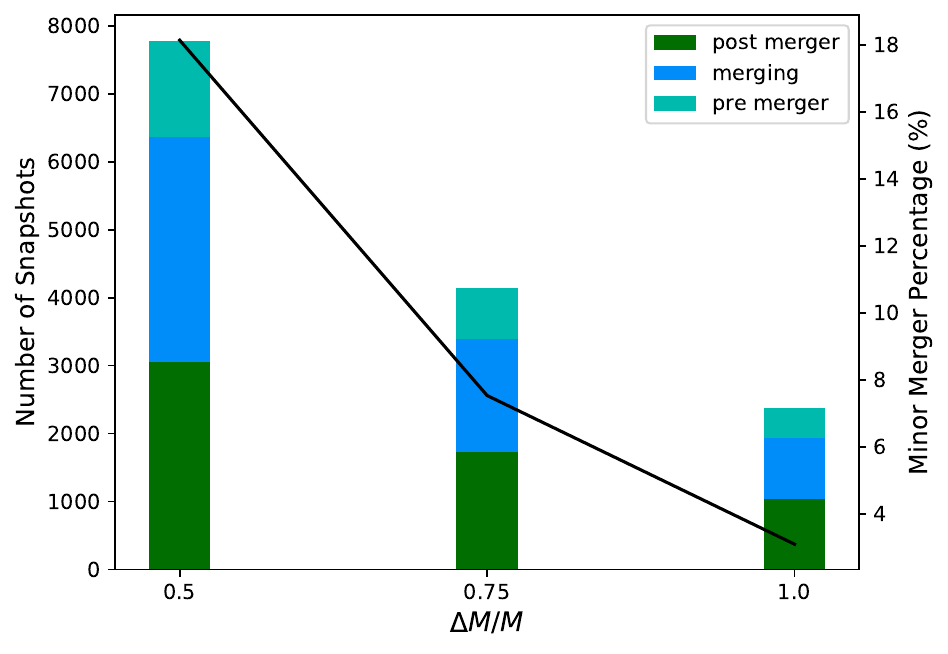}
    \caption{Counts for the number of merging snapshots (left $y$-axis) in \textsc{The Three Hundred} catalogue for three threshold values of $\Delta M / M$ as defined by Equation \ref{eqn:mass_increase}. Counts are split by merger phase, with post- and currently merging groups dominating over the pre-merging phase. The line represents the percentage of total counts belonging to minor mergers (right $y$-axis).}
    \label{fig:MassThresholdEvaluation}
\end{figure}

\citet{Behroozi_2015} outline inaccuracies for mass recovery in position-space halo finding algorithms, such as the \textsc{ahf} used in this study. Following these results, we exclude samples with mass increase ratios $\Delta M/M \geq 3$ to minimise identifications of unphysical mass growth. Selecting the mass increase threshold to implement on the remaining clusters is a trade-off between including a representative sample of mergers and capturing actual merger events as opposed to general accretion. To decide this threshold, we refer to Fig.~\ref{fig:MassThresholdEvaluation} which shows the number of mergers captured, the percentage of minor and merger populations, as well as the pre-, post- and merging distributions for three discrete values of $\Delta M/M $. We note the increase in the total number of snapshots and the minor merger percentage with decreasing $\Delta M/M $. This led us to adopt a lower threshold than in \citetalias{Contreras2022}, since adding more training data improves the generalization of a machine learning model. In other words, the model is less likely to learn patterns that exist only in the training set and do not apply to unseen data. Although the threshold of $\Delta M/M = 0.5$ would increase the number of training samples when compared to $\Delta M/M = 0.75$, it more than doubles the percentage of minor merger as well as risks including a higher percentage of simple mass accretions. One of the main goals of this model is to definitively identify merging clusters, so in an effort to reduce contaminants, and since the $\Delta M/M = 0.75$ threshold has ample data for training, we adopt this value as the cutoff.

We also note that our merger definition includes all merger phases, Fig.~\ref{fig:MassThresholdEvaluation} shows that there is an uneven distribution of these phases, with fewer clusters in the pre-merging phase compared with post- and current merging phases. This is due to the fact that the average pre-merger phase is shorter than both other phases, therefore fewer halos are found during this transition.

\begin{figure}
	\includegraphics[width=\columnwidth]{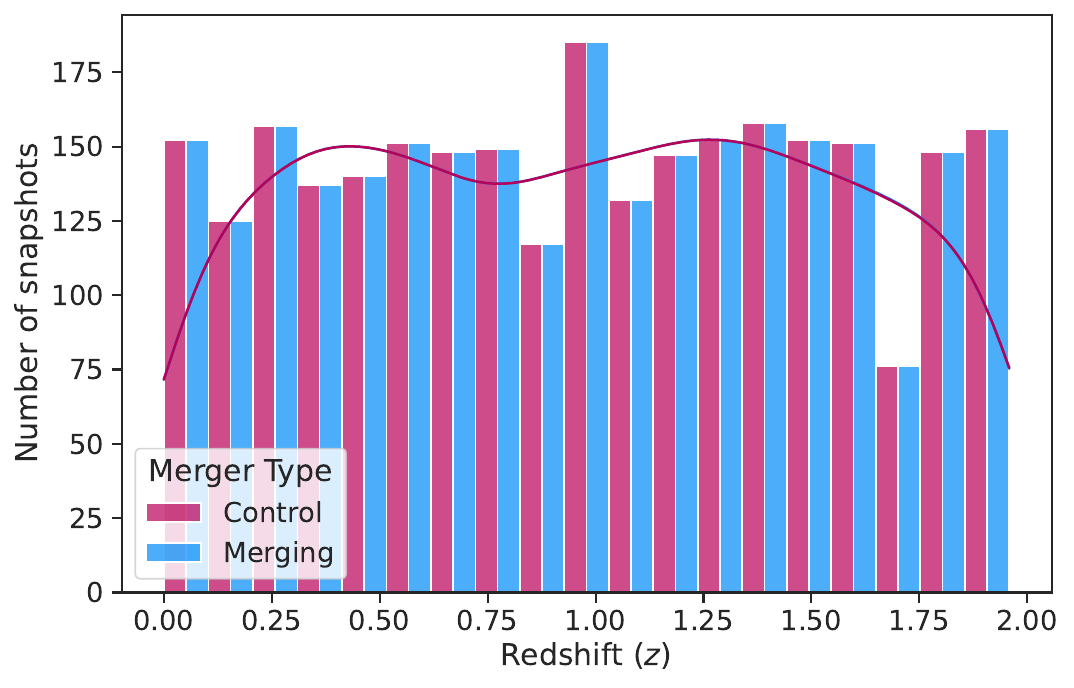}
    \caption{Histogram of snapshots binned by redshift for all of the merging clusters, as well as the control population. Solid lines show the kernel density estimates for each group. A redshift threshold of $z = 2$ has been imposed.}
    \label{fig:redshift_cut}
\end{figure}

The final cut we impose on the merging sample is in redshift-space, requiring $z\leq 2$. This is to constrain the environmental origins of clusters: since we are creating an idealised set of nearby clusters, those at higher redshift are less likely to form in a similar way to those we see nearby. This cut also further reduces any inaccuracies from the \textsc{ahf} halo finder, which are more prevalent at higher redshift. 73\% of the entire simulation's merging cluster population is kept after this cut. Fig.~\ref{fig:redshift_cut} shows the resulting, relatively flat distribution of redshifts. Sampled mergers should therefore have a roughly equal likelihood of coming from any redshift in that range. Anything beyond $z = 2$ is left for future work. If a merger begins before $z = 2$ then only the snapshots taken after this threshold are included as is illustrated in Fig.~\ref{fig:MAH_XDS_evolution} with cluster 158.  


\subsection{Control sample creation}
\label{subsec:ControlSample}

In order to provide a balanced data set for training our binary classification model, a non-merging `control' sample is generated. Regions are selected from the remaining main branch clusters within \textsc{The Three Hundred}'s catalogue, excluding any clusters that meet the $\Delta M / M \geq 0.75$ criteria and exist in either post- or pre-merger state. Selecting from the main branch was necessary to ensure there was no overlap with the merging population, but this limits the diversity of control group clusters. Expanding beyond the main branch is a computationally expensive task and we save it for future research. A cluster's mass has a large influence on its intracluster medium gas and distribution; studies such as \citet{deandres22_cnn} have directly used mock images of Compton $y$-maps to calculate the cluster mass. As such, we want to ensure limited mass bias in our training data, whilst also allowing the mass to vary in case of any causal relationship between the mass and merger probability. We therefore create a cluster control sample with the same redshift distribution as the merging sample. This is achieved by random stratified sampling from the available clusters; see Fig.~\ref{fig:redshift_cut}, which highlights the equivalent distribution. 

\begin{figure}
	\includegraphics[width=\columnwidth]{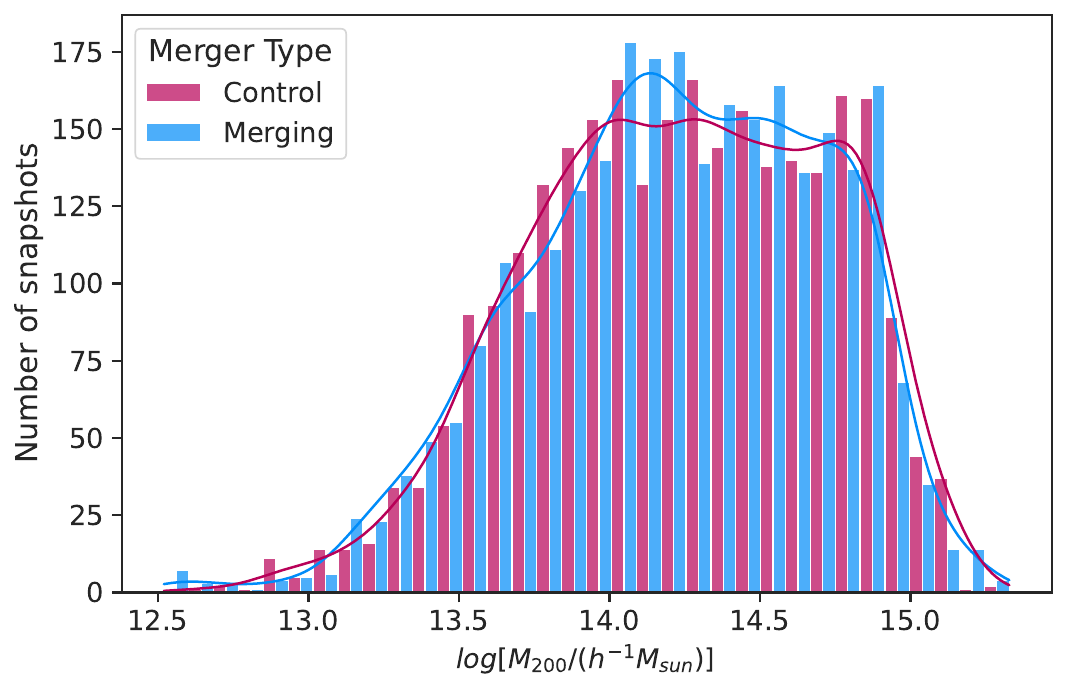}
    \caption{Histogram of snapshots binned by mass for all of the merging clusters, as well as the control population. Solid lines show the kernel density estimates for each group.}
    \label{fig:control_merger_mass_distribution}
\end{figure}

We then investigate the resulting mass distribution for these samples as seen in Fig.~\ref{fig:control_merger_mass_distribution}. Again, the control population closely follows the mass distribution of the total merging population. It is worth noting that within the merger population, the merging phases have different mass distributions. Since the pre-merger population exists before the extra mass from the merger has been accreted, the mass tends to be lower than the total group. Whereas, for the post-merger population, the masses are generally higher as material has already been accreted, and these exist at a lower redshift than the merging population. 


We now have our full training set as described consisting of 5,478 total snapshots, which translates to 158,862 total projections for training the machine learning model. The merging and control groups have the same number of images in each group, same median redshift of 0.99 and comparable median masses of 1.7 and 1.8 $\cdot 10^\mathrm{14} h^\mathrm{-1}$ \(\textup{M}_\odot\) respectively. They do, however, differ in the number of unique regions within each group, where the control is sampled from the full 324 simulated regions, and the mergers only occur in 208. 


\section{Machine learning methods}
\label{sec:Methods}
Before the images are fed into a model they need to be pre-processed and optimised for the algorithm. Firstly, the images are resized using the \textsc{OpenCV} \citep{opencv_library} python package; each 640 x 640px$^{2}$ image is interpolated onto a 96 x 96px$^{2}$ image. We perform this image reduction in order to speed up the training process. Visual inspection of the maps shows that the decrease in resolution does not cause significant loss of information in terms of the gas distribution, and testing a model with 200 x 200px$^{2}$ resolution also gave similar performance.

We then split our data into train, validation, and test data sets. The purpose of each set are as follows: 
\begin{itemize}
    \item \textbf{Train set}: used to teach the model patterns from the data;
    \item \textbf{Validation set}: shown to the model at the end of each training epoch to evaluate its performance on unseen data throughout the learning process and adapt hyperparameters accordingly; and
    \item \textbf{Test set}: used to evaluate the model performance on unseen data once the training process is complete. 
\end{itemize}

When determining the data splits it is important to note that this is an imbalanced data problem; only 17\% of the total main branch clusters are classified as merging across the redshift range of $0 \leq z \leq 2$. When building our training and validation sets, we want an even distribution of both classes, which is why we include equal numbers of control and merger populations, under-sampling from the control. We reserve 10\% of our pre-processed images for testing. Of the remaining snapshots, 90\% are then randomly selected for training and 10\% for validation. For the test set, we then randomly select 17\% of the merging class, and all control, to create a set with the same class imbalance that we see in the simulation data. During the random selection process, we ensure that none of the augmented images of the same cluster snapshot appear across different sets. This is to avoid any skew in our results from the model evaluating performance on cluster states that the model has already seen. The number of images and percentage of mergers in each group are displayed in Table \ref{tab:TrainingSets}.

\begin{table}
	\centering
	\caption{Describes the number of images and percentage of positive merger class belonging to the training, validation and test sets.}
	\label{tab:TrainingSets}
	\begin{tabular}{lcc} 
		\hline
		 & Number of Images & Mergers (\%) \\
		\hline
		Train & 128673 & 49.9 \\
        Validation & 14297 & 50.3\\
        Test & 9468 & 17.0\\
		\hline
	\end{tabular}
\end{table}

We normalise each input image such that for each separate channel -- SZ and X-ray -- the pixel values have a mean of 0 and a standard deviation of 1. This is implemented using the \textsc{keras} normalisation pre-processing layer according to Equation \ref{eqn:normalisation},
\begin{equation}
\label{eqn:normalisation}
I^\mathrm{norm}_{i,j}= \frac{I_{i,j} - \mu(I_\mathrm{train})}{\sigma(I_\mathrm{train})}
\end{equation}
where $I^\mathrm{norm}_{i,j}$ is the pixel-wise value for the normalised image,  $I_{i, j}$ is the pixel-wise value of the original image and $I_\mathrm{train}$ represents the array of images in the training set. The purpose of this normalisation is to ensure all input data is on the same scale, stabilising the gradients used in model calculations and helping the model converge quicker. By implementing this normalisation as a layer in the model it can be fed resized, raw data when making model predictions. The validation and test set are normalised according to the distribution of the training set to prevent data leakage.

Choosing the appropriate machine learning tool is a balance of model accuracy and explainability. Deep learning \citep{lecun2015} models build upon the concept of multi-layer perceptrons, with deep layers of fully connected neurons and non-linear activation functions. These characteristics lend to high levels of model complexity and as such they are very powerful at performing both classification and regression tasks. Convolutional neural networks (CNNs, \citealt{Lecun_98}) are a type of deep learning model that are well suited to image data; they apply kernel operators to images to extract important features, before feeding into fully connected layers to learn patterns that are most important for predictions. For a review on deep learning methods, including further details on CNNs, see \citet{schmidhuber_2015}. Various CNN architectures have been developed and fine-tuned to advance performance; see the review in \citet{Khan_2020}.

In this paper we build and test three separate machine learning models with differing inputs. These include SZ only, X-ray only, and a combined 2-channel input containing both SZ and X-ray images. For each of these models, a different architecture was implemented, each based on the VGGNet architecture introduced by \citet{simonyan2015deep}, and later applied to galaxy clusters to predict their mass in \citet{Ntampaka2019} and \citet{ deandres22_cnn}. In all cases, the output of the network is a single neuron, which quantifies probability that the input cluster is merging, and is used to classify the cluster into being a merger or a non-merger. For the 2-channel model we tested a multi-head architecture, with separate convolutional layers feeding into joint fully connected layers, similar to that used by \citet{Yan2020}. This architecture was favoured over a 2-channel single-head architecture due to the relative ease in which contributions from each of the channels can be separated when interpreting the model in a multi-head architecture.

For all three input data scenarios we perform a grid-search over hyperparameter space (outlined in Table \ref{tab:hyperparameters}) to find the best performing model. The model is built using the \textsc{keras} package with a \textsc{Tensorflow} backend; tests are run on a single NVIDIA A100 GPU and use several different evaluation metrics to gain a holistic view of performance. These metrics and the results are outlined in Section \ref{sec:Results}. We find that the best multichannel model outperforms the best single channel models with the following architecture:

\begin{enumerate}
    \item[\textbf{Separate heads for SZ and X-ray:}] 
    \item Normalization layer
    \item $3\times 3$ convolution, 16 filters
    \item $2\times 2$ stride-2 max pooling 
    \item $3\times 3$ convolution, 32 filters
    \item $2\times 2$ stride-2 max pooling 
    \item $3\times 3$ convolution, 64 filters
    \item $2\times 2$ stride-2 max pooling 
    \item Global average pooling 
    \item[]
    \item[\textbf{Combined:}]
    \item Concatenate
    \item 200 neurons, fully connected
    \item 200 neurons, fully connected
    \item 100 neurons, fully connected
    \item 64 neurons, fully connected
    \item 32 neurons, fully connected
    \item Output neuron.
\end{enumerate}

\begin{figure*}
	\includegraphics[width=\textwidth]{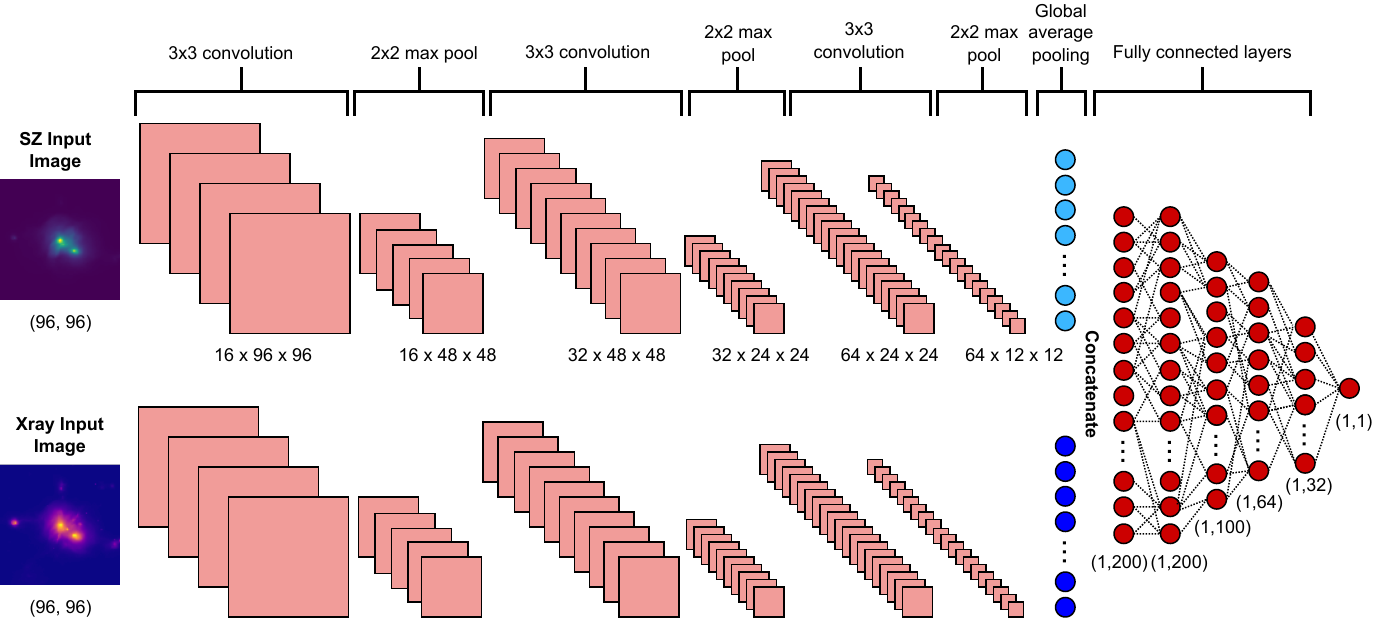}
    \caption{Architecture of the best performing CNN. The top and bottom flows represent each head for the two input channels. There are 3 convolutional blocks for each, consisting of one convolutional layer, and one max pooling layer. Each head then undergoes global average pooling and is concatenated with the other. The fully connected layers are then shown with their corresponding shapes, outputting a single neuron for prediction.}
    \label{fig:CNNArchitectureDiagram}
\end{figure*}

The convolutional layers all have `same' padding, and are followed by a rectified linear unit (ReLU, \citealt{Nair2010RectifiedLU}) activation function. Each pooling layer has `valid' padding. Global average pooling is employed on each head before concatenating into a 1D vector. Each dense fully connected layer is followed by a ReLU activation function. After the first two fully connected layers there is a dropout layer with 20\% of the neurons missed to regularise the model. The final layer has a sigmoid activation to transform the output into a probability score between between 0 and 1. This output neuron is the node that determines the classification of each image. By setting a probability threshold (e.g. 0.5), anything above this threshold will be classified as a merger. This architecture is also presented in Fig.~\ref{fig:CNNArchitectureDiagram}.

We use the Adam Optimiser \citep{kingma2017adam} with a binary cross entropy loss function, optimising for the accuracy on the validation set with a learning rate of $10^{-4}$. The model uses an early stopping function which stops training once the validation loss doesn't improve over a period of 15 epochs. This network took a total of 23 epochs to train averaging 19s per epoch. We perform 10-fold cross validation on this model, keeping the same 10\% as the test set and splitting the remaining training portion into 10 folds. For each iteration, we select a different fold as the validation set and the remaining 9 folds as the train set. 

\begin{table}
	\centering
	\caption{Hyperparameters tested on each model to optimise model performance.}
	\label{tab:hyperparameters}
	\begin{tabular}{lcccc} %
		\hline
        Hyperparameter & Options \\
        \hline
		Learning rate & $10^{-3}$, $10^{-4}$, $10^{-5}$ \\
        Kernel size & (3,3), (5,5) \\
        Number of convolutional layers & 2, 3, 4, 5 \\
        Number of dense layers & 3, 4, 5 \\
        Dropout percentage & 0.2, 0.5\\
		\hline
	\end{tabular}
\end{table}

\section{Results}
\label{sec:Results}
In order to evaluate the success of our models we combine a number of evaluation metrics. As a classification problem we are concerned with the number of predictions that the model gets correct, i.e. the number of true positive (TP) and true negative (TN) outcomes. We are also interested in the number of incorrect predictions, i.e. false positive (FP) and false negative (FN) outcomes, where positive implies a merger prediction. Fig.~\ref{fig:confusion_matrix} is a confusion matrix showing the grouping of predictions for the test data. These are calculated by running the test set through our model and comparing the predictions to actual classifications. The goal of a classification problem is to minimise the FN and FP populations. We combine the results for each of these outcome groups to create the following evaluation metrics:

\begin{figure}
	\includegraphics[width=\columnwidth]{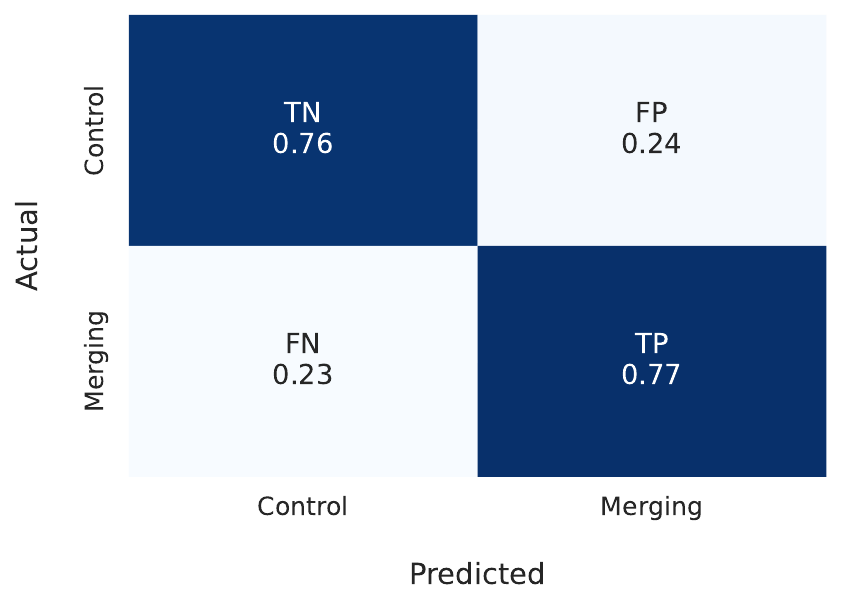}
    \caption{Confusion matrix calculated on the test set for our best model using both SZ and X-ray inputs. The {\it y}-axis represents whether the cluster is actually merging or not, and the {\it x}-axis splits into the model predictions. The values are the number of predictions in each group normalised by the number of elements in each class.}
    \label{fig:confusion_matrix}
\end{figure}

\begin{enumerate}
    \item \textbf{Precision:} true predicted positives out of total positive predictions, $TP / (TP + FP)$, where we define the classification threshold (or probability above which a cluster is classified as merging) as 0.5;
    \item \textbf{Recall:} true predicted positives out of total actual positives, $TP / (TP + FN)$, at a classification threshold of 0.5;
    \item \textbf{ROC-AUC:} area under a receiver operating characteristics curve, which plots the relationship between the true positive rate (TPR, same as recall) and the false positive rate, $FPR = FP / (FP + TN)$, for different classification thresholds. It measures how effectively the model orders predictions in terms of their merger probability. A score of 0.5 is equivalent to a random classifier and 1 is a perfect classifier. In our imbalanced data case, the ROC curve is skewed towards lower values for FPR, biasing the ROC-AUC score towards higher values;
    \item \textbf{PR-AUC:} area under a precision-recall curve. As with ROC-AUC, this metric is independent of classification threshold. It does not take into account the number of TN predictions, as such it is biased towards evaluating the positive class. Values for this vary from 0 to 1; in this case a random classifier would be expected to return a PR-AUC equal to the class imbalance (0.17 on the test set), so values higher than this are an improvement by the model;
    \item \textbf{Balanced Accuracy (BA):} an average of the accuracy of the two classes, as defined by Equation \ref{eqn:balanced_accuracy},
    \begin{equation}
    \label{eqn:balanced_accuracy}
    BA = \frac {1}{2} \left( \frac{TP}{TP + FN} + \frac{TN}{TN + FP}\right).
    \end{equation}
    BA values vary similar to accuracy, varying from 0 to 1. BA is a helpful metric for imbalanced data problems, where the classical accuracy score will perform well even if the model isn't predicting any of the minority class;
    \item \textbf{F$_\mathrm{1}$ Score:} otherwise known as the harmonic mean of the model's precision and recall as described by Equation \ref{eqn:f1_score},
    \begin{equation}
    \label{eqn:f1_score}
    F_\mathrm{1} = 2 \cdot \frac{\mathrm{precision} \cdot \mathrm{recall}}{\mathrm{precision} + \mathrm{recall}}.
    \end{equation}
    The F$_\mathrm{1}$ score is also a useful metric when considering imbalanced data problems, however as with PR-AUC, it does not include TNs.
    
\end{enumerate}

It is important to consider the intended use of the model when evaluating its performance. If we are interested in using the model to curate a sample of merging clusters to study, then we may solely be interested in the ability of the model to identify the positive class, regardless of the control. However, if we are interested in its ability to determine the likelihood of an observed cluster currently undergoing a merger, then we may be more interested in a balanced view of performance that considers both classes, so we would focus on metrics such as BA.  

\begin{figure}
	\includegraphics[width=\columnwidth]{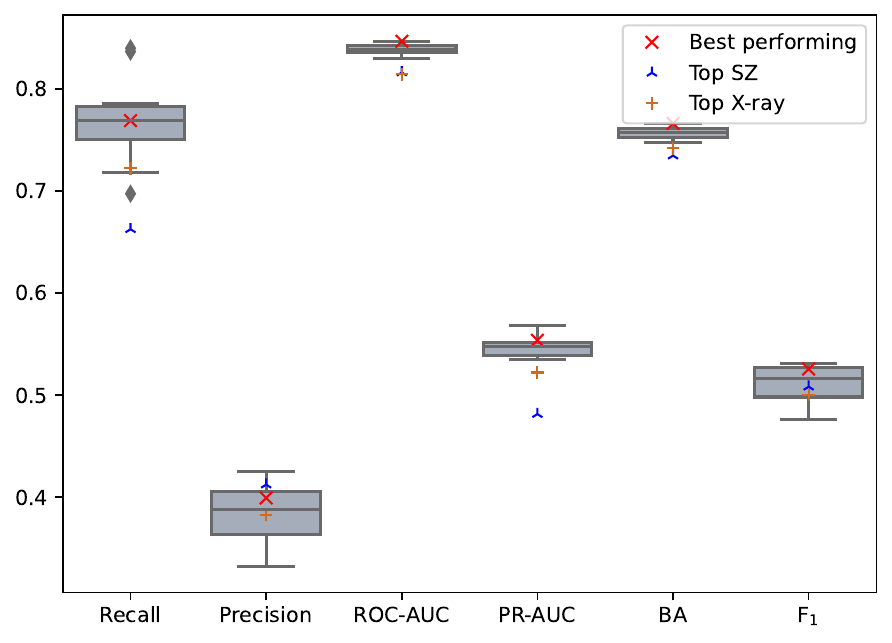}
    \caption{Distribution of evaluation metrics for k-fold cross validation results tested on the hyperparameter-optimised model with both SZ and X-ray inputs. Each box plot represents the distribution of performance for that metric when evaluated on each of the 10 folds. Median values are indicated by the horizontal lines between the bars. Outliers are shown by the grey diamonds. Red crosses indicates our selected best model optimised for BA, blue triangle and orange plus points indicate the corresponding best results for SZ and X-ray only model respectively.}
    \label{fig:evaluation_kfold}
\end{figure}

Fig.~\ref{fig:evaluation_kfold} shows the outcome of the k-fold cross-validation on our hyperparameter-optimised model. Each box plot represents the distribution of performance for that metric when evaluated on each of the 10 folds. We select our best performing model as that with the highest BA measure. This results in a model with ROC-AUC, PR-AUC, BA and F$_{1}$ scores of 0.85, 0.55, 0.77 and 0.53 respectively. In summary, this model does a good job at correctly identifying all mergers in a population, however it is less good at distinguishing them from the control. In other words, the model over-predicts the number of merging clusters, leading to high numbers of FPs, lower precision, and therefore lower PR-AUC and F$_{1}$ scores. The box plots in Fig.~\ref{fig:evaluation_kfold} show little scatter around the evaluation result for each metric, giving us confidence in the reproducibility of results. Higher scatter, and some outliers, in the recall relative to other metrics is due to a low value for the denominator in the metric, so any TP variability drives higher ranges for recall values. The precision values appear low, however, we would expect a random classifier to have a precision equal to the proportion of the merging class. 17\% of our test samples are merging, therefore a median precision of 0.39 is greater than a two-fold improvement on a random classifier.

This best (multiwavelength) model is a 3\% and 6\% improvement on the SZ-only and X-ray-only models in terms of BA and F$_{1}$ score respectively. By varying the probability threshold on this best model from 0.5 to 0.8 (i.e. classifying anything with a probability of ${{>}}0.8$ as a merger), we can achieve a maximum F$_{1}$ score of 0.55. This new threshold increases the precision to 0.55 from 0.4, but comes at the cost of reducing recall to an equivalent value of 0.55 from 0.77.

We adopt standard methodologies used to determine cluster dynamical state to compare our model's performance with conventional methods. The morphological indicators we compare consist of asymmetry, concentration and centroid shift, as defined in \citet{De_Luca_2021}. For the asymmetry we measure within an $2R_{200}$ radius, for the concentration we use concentric circles $0.1R_{500}$ to $R_{500}$, and for the centroid shift we use concentric circles at $R_{200}$ and $2R_{200}$.  We are not aware of existing literature that uses these metrics to classify clusters into merger states. As such, we adopt the thresholds that optimise the balanced accuracy metric on our data. These results can be found for both SZ and X-ray filters in Table ~\ref{tab:Conventional_method}. Our model outperforms all three morphological indicators in both filters.


\begin{table}
	\centering
	\caption{Comparison of our model performance to morphological indicators used to determine cluster dynamical stability. Threshold values for each morphological indicator are determined by optimising for the balanced accuracy measure.}
	\label{tab:Conventional_method}
	\begin{tabular}{llcccc} 
		\hline
		  Model method & Imaging Filter & BA & $F_{1}$ & Recall & Precision \\
		\hline
		Asymmetry & SZ & 0.68 & 0.43 & 0.60 & 0.32 \\
            Asymmetry & X-ray & 0.69 & 0.45 & 0.58 & 0.36 \\
            Centroid Shift & SZ & 0.59 & 0.32 & 0.41 & 0.25 \\
            Centroid Shift & X-ray & 0.59 & 0.33 & 0.53 & 0.24 \\
            Concentration & SZ & 0.60 & 0.33 & 0.49 & 0.25 \\
            Concentration & X-ray & 0.61 & 0.35 & 0.51 & 0.27 \\
            CNN & SZ & 0.73 & 0.51 & 0.66 & 0.41 \\
            CNN & X-ray & 0.74 & 0.50 & 0.72 & 0.38 \\
            CNN & SZ \& X-ray & 0.76 & 0.53 & 0.77 & 0.40 \\

		\hline
	\end{tabular}
\end{table}

\begin{figure}
	\includegraphics[width=\columnwidth]{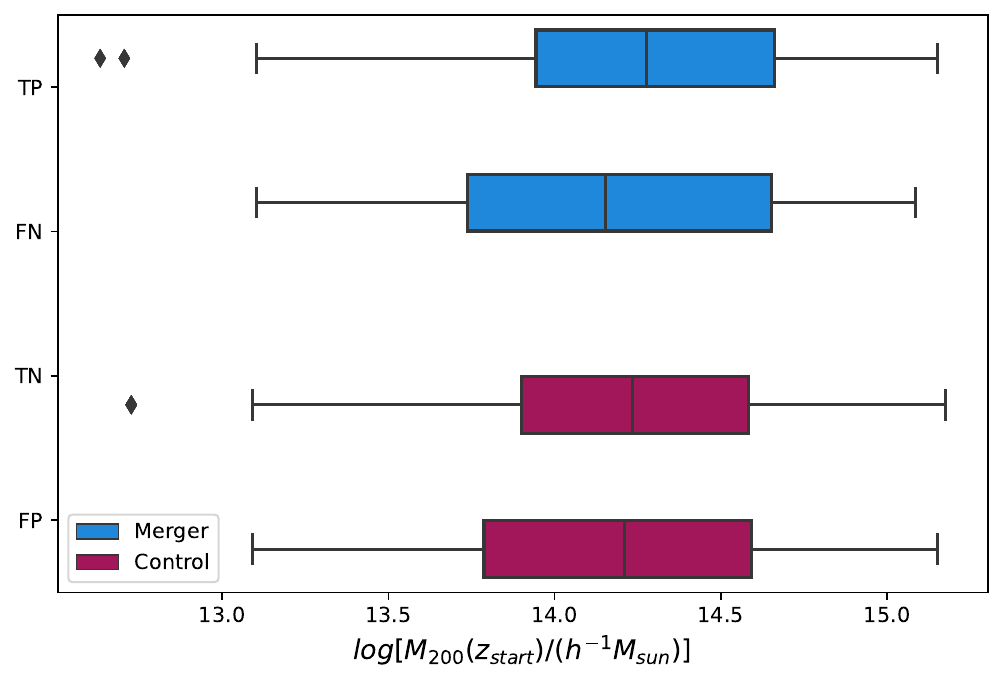}
    \caption{Distribution of mass for each classification group for the model. All groups follow similar distributions, eliminating any mass bias concern.}
    \label{fig:mass_distributions_predicted}
\end{figure}

The mass of a cluster is directly linked to the distribution of intracluster medium gas, hence we can infer the mass using SZ and X-ray observations \citep{Ntampaka2019, deandres22_cnn}. We observe the mass distributions of each prediction group in Fig.~\ref{fig:mass_distributions_predicted}. Since all distributions are similar, we exclude the possibility of the model holding any mass bias.

\begin{figure}
	\includegraphics[width=\columnwidth]{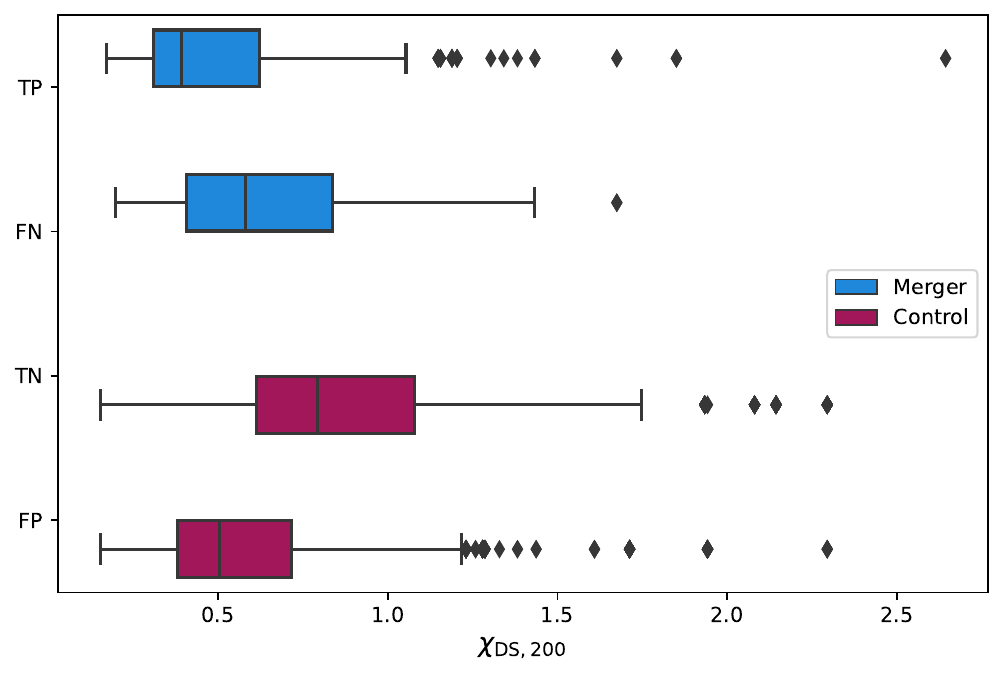}
    \caption{Distribution of the relaxation parameter for each classification group for the model. The correctly predicted merging population tends to contain less-relaxed clusters than correctly predicted non-merging (control) populations. The FN and FP groups have higher and lower respective median $\chi_\mathrm{DS}$ parameters than expected for their class label.}
    \label{fig:XDS_model_classifications}
\end{figure}

We analyse the correlation between the merger probability and relaxation parameter for each cluster snapshot, finding significant ($p = 0$) Pearson and Spearman rank-order correlation coefficients of -0.43 and -0.52 respectively. This implies a negative, slightly non-linear relationship. The negative correlation makes intuitive sense as clusters with lower $\chi_\mathrm{DS}$ values are more likely to be merging. We investigate the distribution of $\chi_\mathrm{DS}$ values for each prediction category, as shown in Fig.~\ref{fig:XDS_model_classifications}. This figure clearly shows that incorrect predictions have different $\chi_\mathrm{DS}$ distributions to their correctly predicted counterparts, and highlights the diversity of cluster merger events. It exhibits a scale of disturbance, whereby the TPs are the most disturbed, followed by FPs, FNs and then finally the TNs. This helps to explain the model performance and gives us more confidence in the probability measure serving as an indicator of disturbance. The TNs have the largest spread of all of the groups. This is probably as they contain stable, relaxed clusters, as well as those that are disturbed but do not meet the mass threshold for a merger.

\begin{figure}
	\includegraphics[width=\columnwidth]{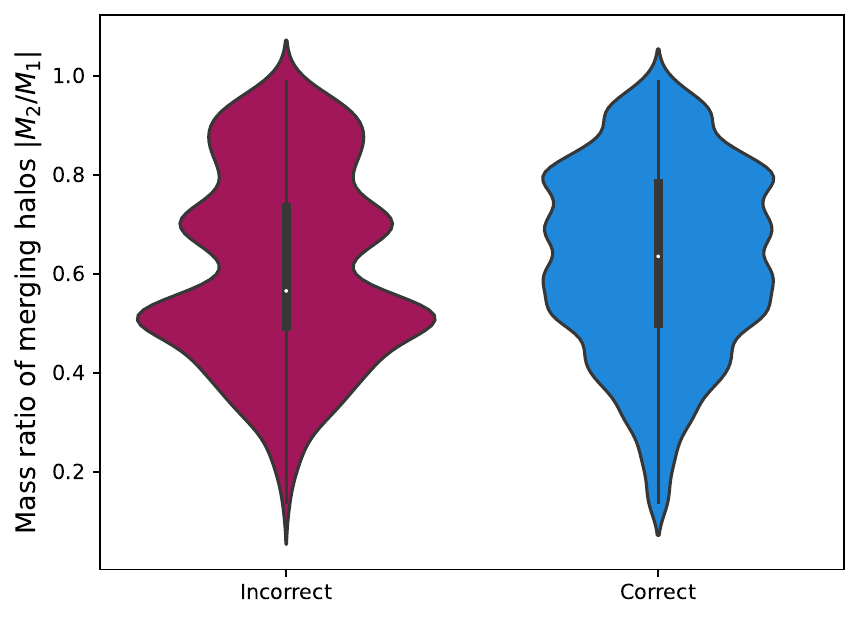}
    \caption{Distribution of the mass ratios of merging halos, comparing incorrect to correct classifications. Violin plot indicates the median with a white dot, the interquartile range is indicated by the thicker black line, and the thinner line shows the remaining distribution, excluding outliers. The shape of the plot follows the kernel density distribution. Incorrect predictions tend to have smaller mass ratios.}
    \label{fig:mass_ratio_mergers}
\end{figure}

\begin{figure}
	\includegraphics[width=\columnwidth]{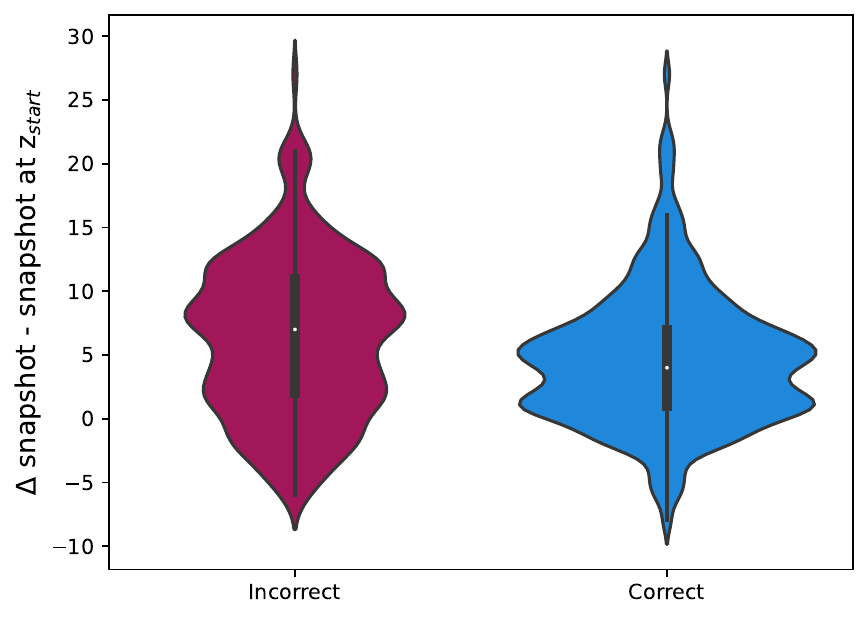}
    \caption{Distribution of the lag in current snapshot values to the snapshot at which the merger begins, comparing incorrect to correct classifications. Violin plot indicates the median with a white dot, the interquartile range is indicated by the thicker black line, and the thinner line shows the remaining distribution, excluding outliers. The shape of the plot follows the kernel density distribution. Incorrect predictions tend to be be further apart in time from the merger event than correct predictions.}
    \label{fig:snap_lag}
\end{figure}

We further investigate the effect dynamical instability has on the predictions for the merging clusters alone in Fig.~\ref{fig:mass_ratio_mergers}. This plot shows the mass ratio of merging subhalos for incorrect predictions compared to the correct predictions. Incorrect predictions tend to have smaller mass ratios, thus are more likely to be minor mergers that cause less disturbance and are more difficult to identify. To substantiate this, we look at the lag between the imaged snapshot and the snapshot at $z_\mathrm{start}$ for the merger. Fig.~\ref{fig:snap_lag} shows that correct predictions tend to be those that are imaged just after the mass threshold is breached and the merger event begins. Whereas, incorrect predictions are more spread out in time, with greater populations just before the merger begins, and a median lag of 7 snapshots after the merger event. This indicates that incorrect predictions tend to occur when the merging cluster is not at the peak of merging activity.

\begin{figure}
	\includegraphics[width=\columnwidth]{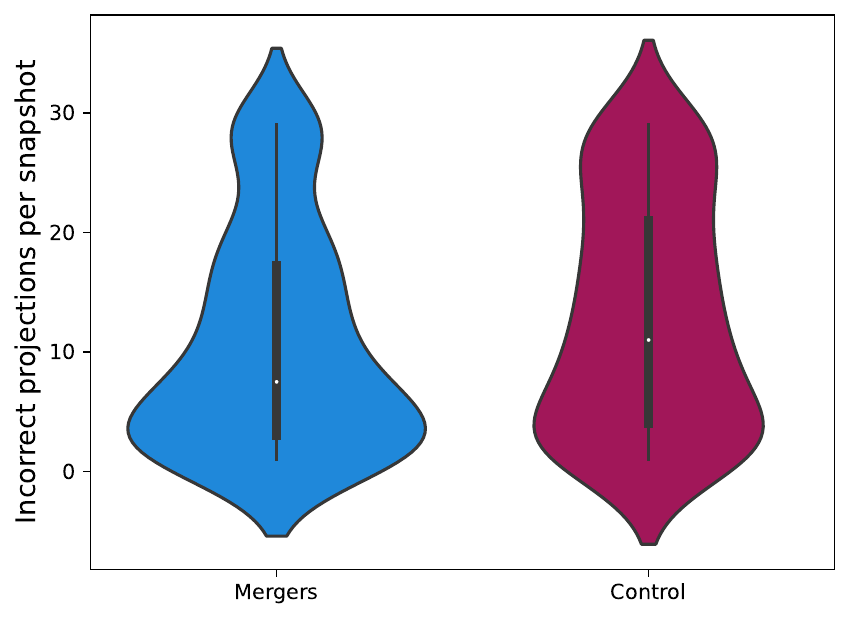}
    \caption{Distribution of the number of projections classified incorrectly per individual snapshot in the test data. Violin plot indicates the median with a white dot, the interquartile range is indicated by the thicker black line, and the thinner line shows the remaining distribution, excluding outliers. The shape of the plot follows the kernel density distribution. Mergers tend to have much fewer incorrect projections per snapshot than non-mergers.}
    \label{fig:incorrect_proj_per_snapshot}
\end{figure}

\begin{figure}
	\includegraphics[width=\columnwidth]{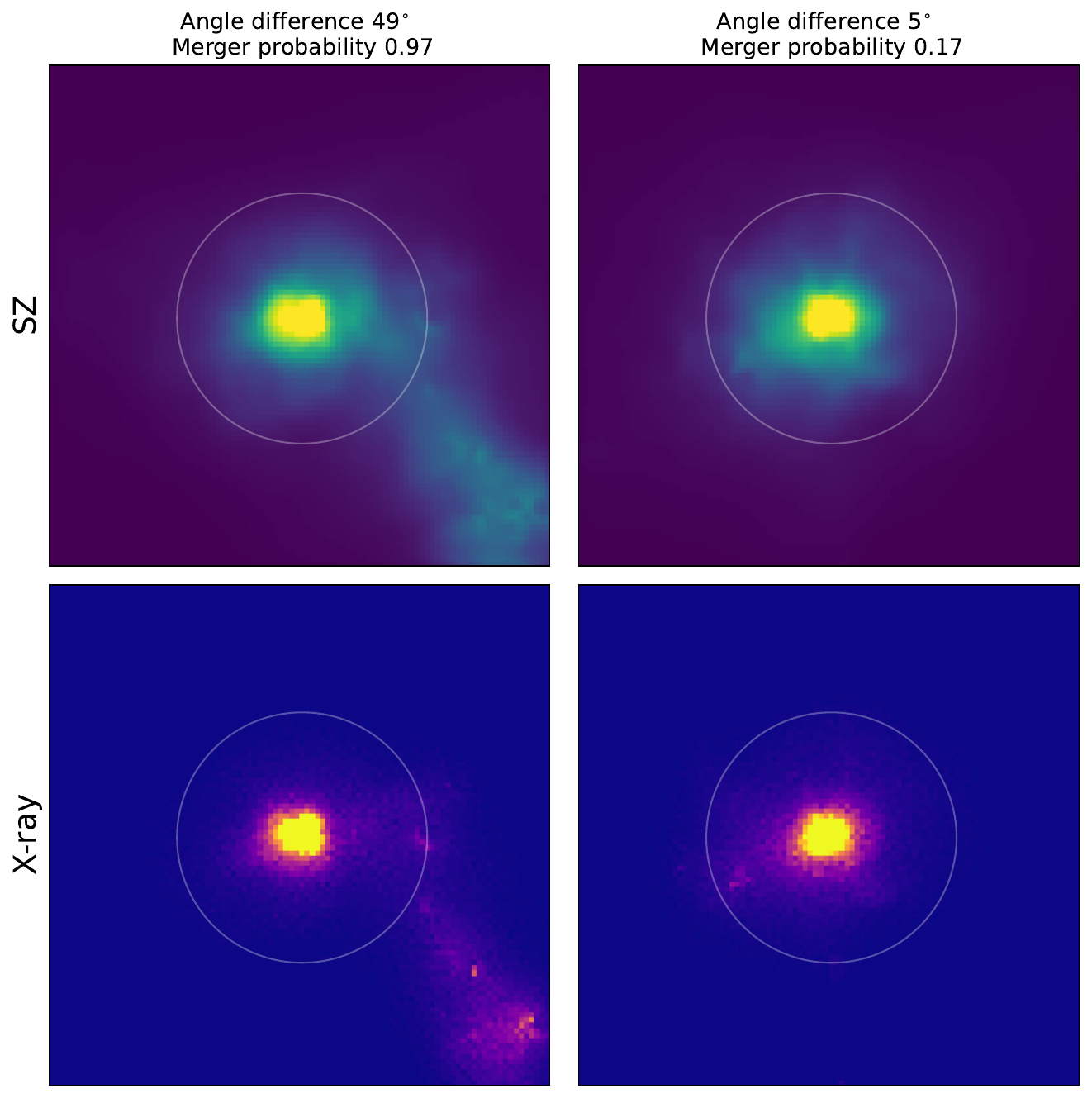}
    \caption{Same cluster imaged along two separate projection axes, indicated by the columns. The left hand projection is separated from the merger axis by 49$^\circ$ and the model assigned a merger probability of 0.97, whereas the right hand projection is separated by 5$^\circ$ and has a merger probability of 0.17. The top pane comprises the SZ signal, the bottom is X-ray. White circle represents the $R_{200}$ radius.}
    \label{fig:projection_example}
\end{figure}

Another important consideration is the effect of the projection axis on the classification, which we investigate by measuring the number of incorrect projections per snapshot as per Fig.~\ref{fig:incorrect_proj_per_snapshot}. The merging group has a median of 7.5 misclassified projections per each snapshot of a cluster, compared with a higher 11 clusters for the control group. This could suggest that merger misclassifications occur when the merger axis aligns with our line of sight. To investigate this further, we obtain the angle between the merger axis, calculated as the vector between the two merging halos at $z_\mathrm{start}$, and the projection axis. We find a number of cases where the misclassified projections for a given cluster align more closely with the merging axis than the correctly classified projections. Fig.~\ref{fig:projection_example} shows an example of a cluster where the projection effect is masking the merger and leading to misclassification. The projection on the right is closely aligned to the merger axis and therefore the interacting gas from the merging halo is hidden along this axis. It assigns the projection a merger probability of 0.17, and therefore misclassifies the cluster. On the other hand, the projection on the left pane has clear separation of the merging bodies along the line of sight, therefore the model correctly assigns it a high probability of merging.
The likelihood of a non-merger event being misclassified is higher, yet still only 22\% of control group snapshots have had ${>}50\%$ of their projections misclassified. By providing explanation into misclassifications, Fig.~\ref{fig:mass_distributions_predicted} to Fig.~\ref{fig:projection_example} provide evidence that the errors that we see within the model make sense intuitively, giving us confidence in the probability estimates.

\section{Interpretation and Discussion}
\label{sec:Discussion}


\subsection{Interpretability}
Deep learning models are famously `black boxes' with multiple layers of non-linear transformations making them difficult to interpret. Methods are continually being developed that help to conceptualise how and why complex models make their predictions, albeit not without challenges as described in a review by \citet{molnar2022}. In order to make sense of the regions in an image that are important to our CNN we adopt the vanilla saliency approach \citep{Simonyan14a}. This method calculates the pixel-wise gradient of the output with respect to the input image, resulting in a map whose pixel values indicate its relative importance in the prediction. We note that the vanilla gradient method is known to suffer from saturation problems among others \citep{Shrikumar2017}, but has been insightful when applied to galaxy clusters, e.g. in \citet{deandres22_cnn}.

\begin{figure}
	\includegraphics[width=\columnwidth]{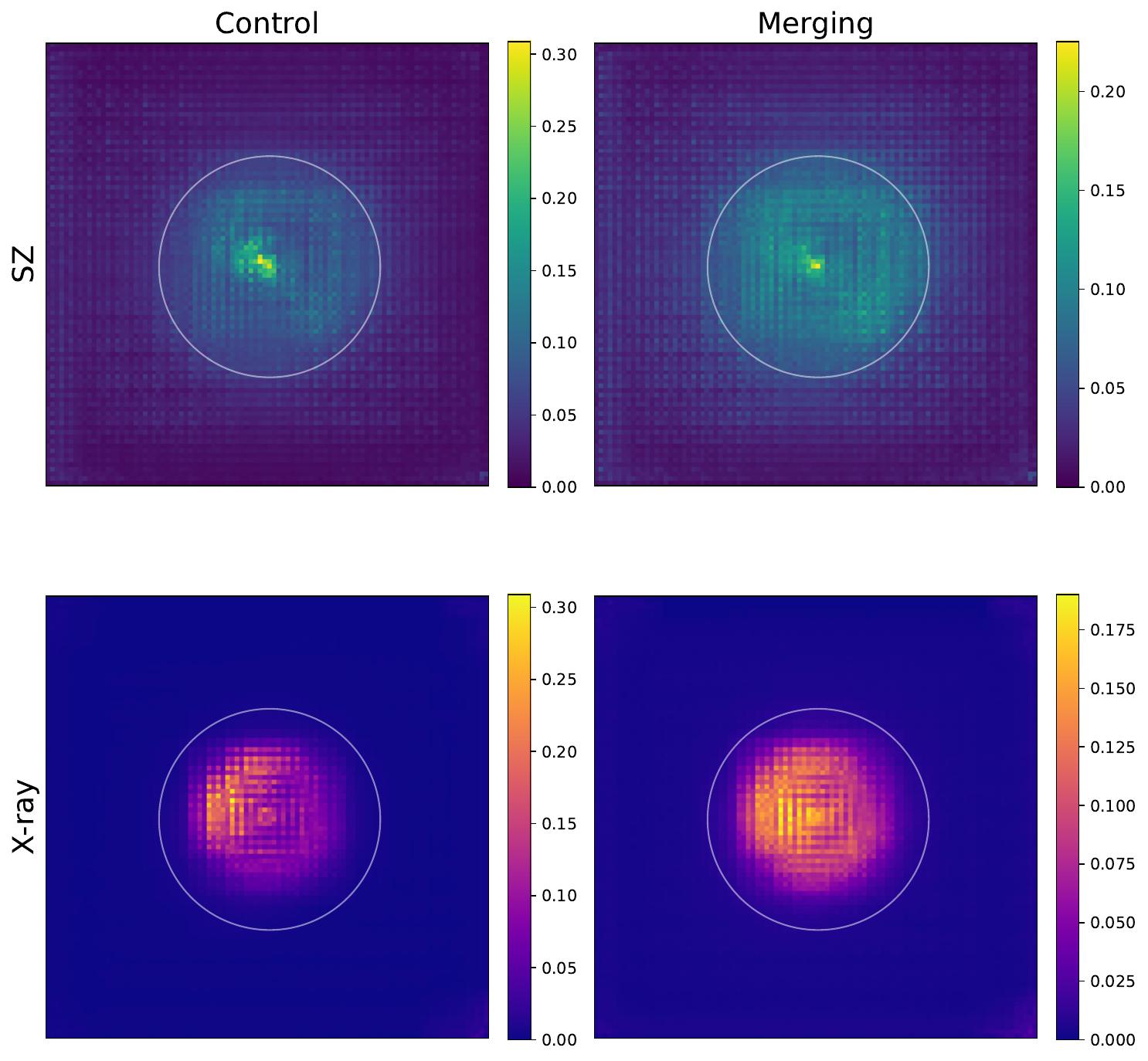}
    \caption{Median pixel values for stacked normalised saliency maps for correctly classified predictions, split by classification label and input data channel. White circles represents the $R_{200}$ radius.}
    \label{fig:stacked_saliency_maps}
\end{figure}

Since this is a classification problem, to compute the gradient we first modify the activation of the final layer of the network from a sigmoid to a linear function. We then back-propagate over the network to produce a saliency map with the same dimensions as the input with values normalised to between 0 and 1. These maps are stacked for each population with their medians calculated at each pixel location as shown in Fig.~\ref{fig:stacked_saliency_maps}, normalised to the maximum of each individual image. Across all groups, the central regions of the cluster are of higher importance to the model compared to the cluster outskirts. This saliency is more concentrated for the X-ray inputs than the SZ channel inputs, out to radii of $\sim0.7 R_{200}$ 
and $\sim1.2 R_{200}$ respectively. This discrepancy is justified by differing scaling of each observation with density ($n$), whereby the Compton-{\it y} parameter scales linearly ($\propto n$) and X-ray flux varies as the square ($\propto n^{2}$) \citep{Mroczkowski2018}. Whilst the distributions for the merging and control groups are relatively similar, we can see that the control group is more centrally concentrated. The fact that the model puts more emphasis on pixels at a larger distance from the cluster centre for the merger population can be explained by the existence of substructures offset from the cluster centre. The logical explanation behind the saliency interpretation gives us confidence that the model is learning physical patterns in the data.

\begin{figure}
	\includegraphics[width=\columnwidth]{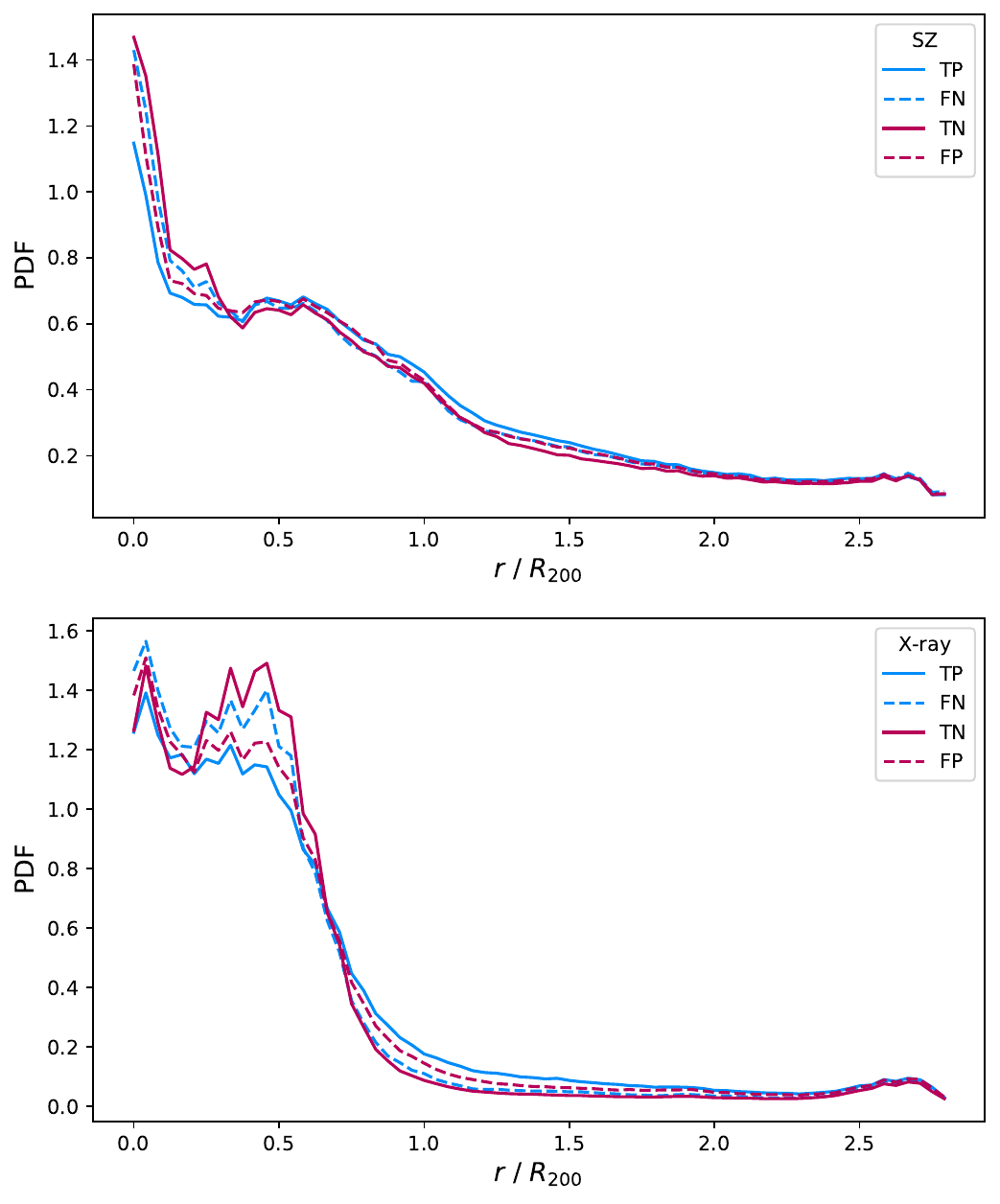}
    \caption{Probability distribution function for radially-binned saliency values of stacked, normalised saliency maps produced using the best, combined SZ \& X-ray model. Top panel: SZ channel input, Bottom panel: X-ray channel input. Blue lines represent the actual mergers, pink is the actual control group, with dashed lines representing the misclassifications for each group. }
    \label{fig:radial_saliency_plot}
\end{figure}

We then bin the saliency maps radially from the cluster centres and calculate the probability density function across the extent of the radial profile of the clusters, as shown in Fig.~\ref{fig:radial_saliency_plot}. The plots reflect the same disparity in the concentrated importance for X-ray vs SZ observations. We also see that across both channels, the incorrect predictions tend towards the distributions for the correct predictions of the opposite group. This is most evident at central radii (SZ: ${<} 0.4 R_{200}$, X-ray: ${<} 0.6R_{200}$), where the FP (FN) group holds more (less) significance than the TN (TP) group. As well as at intermediate radii (SZ: $0.6R_{200}< r < 1.6R_{200}$, X-ray: $0.6R_{200}< r < 2.0R_{200}$), where the opposite is true. Incorrect predictions therefore represent populations that are in some way closer to the opposite classification group, as discussed in Section \ref{sec:Results}. 

There is some asymmetry in the saliency maps shown in Fig.~\ref{fig:stacked_saliency_maps}, suggesting the model is learning some bias in the clusters. From stacking the original training images, there is no evidence of clustering in any given direction from the cluster centre, so this is unlikely to be an issue of bias in the training set. Instead, it may be a product of the CNNs internal processes, or a limitation of the saliency method. For more information on limitations of saliency we refer to \citet{kindermans2017} and suggest that future work could implement practical tests outlined therein. 

\subsection{Limitations \& future work}
The most significant limitation of our model when considering its application to real world observations is the idealised nature of the training data. The images capture more clusters at a better angular resolution than you may expect from an observed sample, and are devoid of instrumental effects or contaminating sources. Future model iterations could build in redshift as an input to the dense layers of the model in a similar architecture to that seen in \citet{Krippendorf2023}. Instrument specifications could be included for both SZ and X-ray observations to set beam smoothing and instrumental noise as well as point-source contaminants.

\begin{figure}
    \includegraphics[width=\columnwidth]{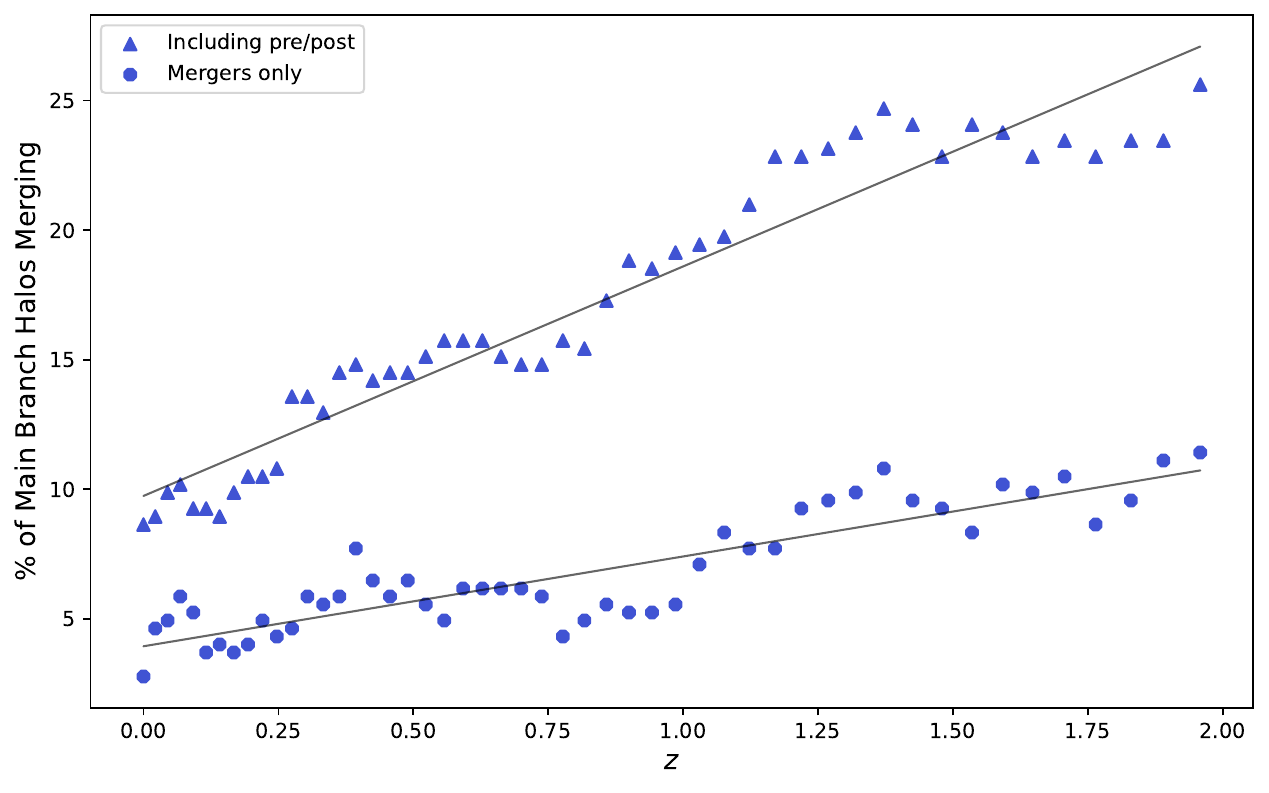}
    \caption{Percentage of the simulation clusters which are merging as a function of redshift. This is split into two groups, with triangles representing the proportion when pre- and post- phases are included in the merger definition, and circles representing the proportion clusters only in the $z_\mathrm{start}$ to $z_\mathrm{end}$ phase.}
    \label{fig:pmerg_redshift_distn}
\end{figure}

When building our test set we set a fixed value for the class imbalance, i.e. the percentage of mergers present in the sample. In reality, the number of mergers occurring at any point in the simulation data on the main branch has a redshift dependence. Fig.~\ref{fig:pmerg_redshift_distn} highlights this relationship, with two separate lines for the case where we have mergers only, and mergers including pre- and post-merger phase. We also note that there is a higher class imbalance if we were to treat the positive class as the merging phase alone. To improve the imbalance problem and capture a wider range of merging states we included all merging phases in this model. There are many open questions surrounding energy dissipation mechanisms within merging clusters, as such the ability to identify merger phase is desirable to link ongoing processes with merger evolution. Building a multi-class prediction model would enable the easy identification of candidates that exhibit strong indicative features of each merger phase. By selecting and investigating example systems for each phase we could perform more comprehensive studies, such as that by \citet{Wilber2019}, understanding their physical properties and gaining insights into the underlying processes transmitting energy. The model may also pick out key features within the intracluster medium gas that distinguish between each merger phase that theory had not yet considered.

In this study we considered SZ and X-ray as two proven effective tracers of the intracluster medium gas within galaxy clusters that are shown to complement each other. There are also other measurements that probe the clusters and could be used to detect merging systems. For example, \citet{Lokas2023} uses cluster temperature maps from simulation data to pick out distinctive bow shocks in the intracluster medium gas. A recent study by \citet{Yuan_2023} suggested that after a merger the cluster gas reaches a relaxed state sooner than the individual member galaxies. Therefore, adding information on the dynamics of the member galaxies could capture a fuller picture of the merger events. This would be equivalent to reframing the prediction outcome of similar such studies that aim to estimate cluster mass \citep{Yan2020, Ho2023}. This would require spectroscopic studies of the cluster to obtain relative velocities, as well as sufficient resolution to distinguish member galaxies within the gas. There are fewer clusters for which this information is available so there would be a more limited application of such a model.

Previous studies have attempted to describe the dynamical state of clusters based on morphological indicators, and in some case classify their merger state as in \citet{parekh2015}. It would be insightful to adapt and apply this model to the same datasets as these studies and compare whether the model determines the same groups as likely mergers. 

We have formed our problem as a binary classification and, whilst acknowledging that the output probability can take a range of values between 0 and 1, this can be restrictive when considering the range of merger situations that can occur. Another possibility would be to form a regression task where the target is the mass increase we defined in Equation \ref{eqn:mass_increase}. Since the mass increase is defined for all snapshots and phases of the merger cycle, within our dataset this would amount to limited variety in the regression values. However, with enough data this could be an interesting future task.

Multiple merger events are not distinguished within our sample, and some are removed due to the $|M_\mathrm{2}/M_\mathrm{1}|\leq 0.1$ threshold. These may have different morphologies and effects on the intracluster medium gas to the binary merger events, and therefore lead to incorrect predictions. With enough data, this could be reframed as a multi-class classification predicting the number of merging systems. We follow mergers only from the main branch halo (Section \ref{sec:Identifying Mergers}), introducing an evolutionary selection bias towards the most massive clusters at low redshift. This is not deemed a significant limitation as this bias is also present in observation. However, by including more halos, we can reduce this bias and increase the training sample size, creating more data to trial multi-class and regression analysis. 

\section{Conclusions}
\label{sec:Conclusion}
We have built a model to identify merging galaxy clusters based on data from 324 clusters within \textsc{The Three Hundred} simulation. The clusters imaged were void of any observational effects and noise from other sources. The model was able to predict the likelihood of a merger event with a BA score of 0.77, $F_{1}$ score of 0.53, PR-AUC of 0.55 and ROC-AUC score of 0.85. By investigating false predictions within our model, we find that missed mergers tend to either be those that are less dynamically disturbed, or are hidden due to the merger axis aligning with our line of sight. For the control population our mass increase threshold for the merger event necessitates a distinction between relaxed and merging events, whereas, in reality we see a spectrum of merger scenarios with varying levels of disturbance. We highlight that the output probability is significantly correlated to the relaxation parameter, such that using the probability should at least indicate the dynamical state of the cluster. 

We sense-check our model using the vanilla saliency method, finding that the model puts more importance on regions at larger radii from the cluster centre when identifying mergers than relaxed. This relation is inversed when looking at false predictions, where incorrect predictions highlight important regions at a greater distance from the cluster centre for the control group, and a closer range for the mergers. The sphere of importance was greater for SZ compared with X-ray, extending to $\sim1.2 R_{200}$ and $\sim0.7 R_{200}$ respectively.

With these considerations in mind, the study demonstrates a successful proof of concept for the application of deep learning in identifying cluster mergers. This model can be used in the immediate future on simulated data as a quick method to determine merger probability without needing to search the clusters mass accretion history. By extending the model to include noise and instrumental effects in the training data, it can be applied to low redshift survey data, speeding up the classification process and reducing the need for visual inspection as in studies such as \citet{Mann_2012}. In future, the scales and redshifts at which these clusters are observed could also be varied. As the number of detectable merging systems increases with upcoming surveys, this model could save a lot of time in quickly identify merging systems.

\section*{Acknowledgements}
YCP and ARA are funded by a Rutherford Discovery Fellowship from the Royal Society of New Zealand. WC is supported by the STFC AGP Grant ST/V000594/1 and the Atracci\'{o}n de Talento Contract no. 2020-T1/TIC-19882 granted by the Comunidad de Madrid in Spain.
D.d.A.,  and  W.C. would like to thank the Ministerio de Ciencia e Innovación (Spain) for financial support under Project grant PID2021-122603NB-C21 and ERC: HORIZON-TMA-MSCA-SE for supporting the LACEGAL-III project with grant number 101086388. This work has been made possible by the `The Three Hundred’ collaboration\footnote{https://www.the300-project.org}. The computations were performed on the R\=apoi high performance computing facility of Victoria University of Wellington. 
The authors acknowledge The Red Espa\~{n}ola de Supercomputaci\'{o}n for granting computing time for running the hydrodynamical simulations of \textsc{The Three Hundred} galaxy cluster project in the Marenostrum supercomputer at the Barcelona Super-computing Center.
We would also like to thank Federico De Luca for sharing code to calculate morphological indicators which aided greatly in understanding the performance of our model.

\section*{Data Availability}

The results shown in this work use data from The Three Hundred galaxy clusters sample. These data are available on request following the guidelines of The Three Hundred collaboration, at \hyperlink{https://www.the300-project.org}{https://www.the300-project.org}. The data specifically shown in this paper will be shared upon request to the authors.

\section*{Code Availability}
The trained model with an example prediction can be found at \hyperlink{https://github.com/ashleigharendt/ClusterMergers}{https://github.com/ashleigharendt/ClusterMergers}.



\bibliographystyle{mnras}
\bibliography{ref_gc23} 




\appendix


\bsp	
\label{lastpage}
\end{document}